\documentclass[aps,prapplied,reprint,superscriptaddress,longbibliography]{revtex4-2}
\usepackage[utf8]{inputenc}
\usepackage{bm}
\usepackage{graphicx}
\usepackage[caption=false]{subfig}
\usepackage[colorlinks=true, citecolor=blue, linkcolor=blue, urlcolor=blue]{hyperref}
\usepackage{pinlabel} 
\usepackage{amsmath}
\usepackage{dcolumn}
\usepackage{amssymb}
\usepackage{natbib}
\usepackage{mathrsfs}
\usepackage[T1]{fontenc}
\usepackage{array}
\usepackage{multirow}
\usepackage{float}

\newcommand{\RN}[1]{%
  \textup{\uppercase\expandafter{\romannumeral#1}}%
}

\begin{document}

\title{Lattice Hamiltonians and Stray Interactions within Quantum Processors}

 \author{Xuexin Xu}
 \email{x.xu@fz-juelich.de}
 \affiliation{Peter Gr\"unberg Institute, PGI-2, Forschungszentrum J\"ulich, J\"ulich 52428, Germany}
 
 \author{Manabputra}
 \affiliation{Department of Physics, Syracuse University, Syracuse, New York 13244, USA}
 
   \author{Chlo\'e Vignes}
 \affiliation{Peter Gr\"unberg Institute, PGI-2, Forschungszentrum J\"ulich, J\"ulich 52428, Germany}
 \affiliation{Department of Electrical Engineering and Computer Science, Massachusetts Institute of Technology, Cambridge, Massachusetts 02139, USA}

 \author{Mohammad H. Ansari}
 \affiliation{Peter Gr\"unberg Institute, PGI-2, Forschungszentrum J\"ulich, J\"ulich 52428, Germany}

\author{John M. Martinis}
 \affiliation{Department of Physics, University of California, Santa Barbara, Santa Barbara, CA, 93106, USA}
 \affiliation{Qolab, Inc, 9854 National Blvd, 1436, Los Angeles, CA 90034, USA}

\begin{abstract}
Developing Hamiltonian models for quantum processors with many qubits on the same chip is crucial for advancing quantum computing technologies. Stray couplings between qubits lead to errors in gate operations. This study underscores the importance of incorporating lattice Hamiltonians into quantum circuit design. By comparing many-body effects with two-body stray couplings, we show how adjusting circuit parameters can enhance two-qubit gate fidelity. We find that loosely decoupled qubits result in weaker stray interactions and higher gate fidelity, challenging conventional assumptions. We investigate the scenario where three-body  $ZZZ$  interaction surpasses two-body  $ZZ$  interactions, highlighting the transformative potential of lattice Hamiltonians for novel multi-qubit gates. Moreover, we investigate the cross-resonance gate within the lattice Hamiltonian framework and examine the impact of microwave pulses on stray coupling. This emphasizes the necessity of developing a comprehensive theoretical framework that includes lattice interactions, which are now critical given the sophistication of contemporary quantum hardware. These insights are vital for developing fault-tolerant quantum computing and next-generation quantum processors.

\end{abstract}
\maketitle

\section{Introduction}
A significant number of qubits and their precise control are essential for practical quantum computations to be feasible. A major challenge in increasing the number of qubits in a quantum processor involves reducing the interference caused by stray couplings from neighboring qubits. This noise interferes with gate operations, often resulting in errors surpassing the acceptable error correction thresholds \cite{preskill2018quantum, bharti2022noisy, gonzalez-garcia2022error}. These inaccuracies primarily stem from the difficulties in achieving both rapid and accurate gates, exacerbated by unwanted interactions between qubits \cite{malekakhlagh2020first-principles, sundaresan2020reducing, cai2021impact}.

One and two-qubit gates are foundational for universal quantum computation, and several strategies have been developed to improve their performance~\cite{hertzberg21laser-annealing,tripathi22suppression,morvan22optimizing,klimov24optimizing}. However, they are not the only gates available \cite{nielsen2002quantum}. Multi-qubit interactions, such as Toffoli gates, together with other gates, can make a different universal set \cite{Georgescu2014quantum}. Although multi-qubit gates do not scale well and may incur additional overhead when constructing many target algorithms, their primary benefit lies in the ability to entangle more than two qubits simultaneously. This capability, in contrast to using a sequence of one- and two-qubit gates, can accelerate computation, reduce circuit depth, and ensure compatibility with state-of-the-art hardware, such as the implementation of multi-qubit interactions using programmable algorithms or in a qudit system. \cite{chu22scalable,fang23realization,nguyen24programmable, nikolaeva24scalable,nguyen24empowering, kim2022high-fidelity, baker2022single, glaser2023controlled, acharya2023suppressing}. If low-error multi-qubit gates can be made and executed, several areas of research will be positively impacted, such as quantum chemistry \cite{cao2019quantum} and machine learning \cite{biamonte2017quantum, pazem2023error}. Beyond the interesting prospect of taming multi-qubit interactions as novel gates, one can acknowledge that such interactions can have detrimental role on two-qubit gates. The effect of multiqubit interactions on two-qubit gate performance, whether subtle or significant, is yet to be fully determined.

In the context of quantum computing, the performance of two-qubit gates is primarily modeled by circuit Quantum ElectroDynamics (cQED) theory, which tends to overlook the effects of multi-qubit interactions, suggesting that their influence is only pronounced at higher orders \cite{ciani2023lecture}.  Several studies have explored the role of multi-qubit interactions in quantum systems \cite{reagor2018demonstration,menke2022demonstration,katz2023demonstration}. Many-body systems exhibit a propensity for long-range correlations, indicating that qubit connectivity enhances delocalization. Recent findings demonstrate the potential for chaos in many-qubit processors within certain parameters, prompting consideration of many-body localization (MBL) theory as a crucial tool \cite{serbyn2013local,huse2014phenomenology,schreiber2015observation}. However, this approach yet underestimates the true strength of multi-qubit interactions, highlighting the need for more sophisticated theoretical frameworks. Understanding and effectively managing these interactions is crucial for the advancement of quantum processors \cite{berke2022transmon, lu2022multipartite}.

In this study, we demonstrate that the presumed hierarchy of many-body interactions is not valid, even within the parameters of quantum computation. Our findings indicate that multi-qubit interactions can be as significant as two-qubit interactions. Further exploration of multi-qubit interactions is essential to understand the adverse effects of three-body entanglement on pairwise interactions.  The paper is organized as follows: First, we study multi-qubit interactions in a qubit lattice, particularly emphasizing the importance of considering the three-body $ZZZ$ interaction. In Sec.~\ref{sec:two1}, we validate our model by evaluating $ZZZ$ in an actual experimental setup in a three-qubit lattice. In Sec.~\ref{sec:two2} we generalize our non-perturbative approach in a triangular lattice of three qubits with three couplers, and highlight scenarios in circuit parameters where attempts to reduce two-qubit gate errors may unintentionally amplify three-body interactions, thereby posing a challenge in enhancing gate fidelity. In Sec.~\ref{sec:three}, we identify specific conditions under which three-body interactions prevail over two-body interactions. Finally we study cases of microwave-driven gates in practical devices and show they can largely benefit from tuning circuit parameters by reducing nonlocal parasitics.

\section{Lattice Hamiltonian}

In quantum processors, qubits are positioned at the vertices of an underlying spatial lattice and interact by intentional couplings along the lattice's edges. Typically, the time variable is treated as continuous. The Hamiltonian of the processor acts as the generator of time evolution of quantum state of the lattice, which is usually derived from the cQED formulation, involving a transformation of the processor's circuit Lagrangian. 

A logical qubit comprises multiple physical qubits, typically arranged at the vertices of a lattice and interacting with their nearest neighbors \cite{fowler2012surface,wu2019sterngerlach,erhard2021entangling}. This arrangement is predicated on the assumption that more distant qubits interact weakly. Yet, in practice, distant qubits may exhibit significant interactions. For instance, in superconducting processors, distant qubits may interact via direct capacitance or inductance. Moreover, weakly-interacting distant qubits with nearly matching frequencies can significantly influence each other, where activating one qubit may lead to the excitation of another.

Modeling in cQED has traditionally centered around pairwise qubit interactions, often overlooking interactions between distant qubits in the effective circuit Hamiltonians.

In 2D lattices, the fundamental building block can be a triangle, geometrically a 2-simplex. For 3D lattices, the basic unit is a tetrahedron, or a 3-simplex, and so forth. A 1D chain, for example, can be partitioned into line segments, which are 1-simplices. Thus, in cQED, pairwise interactions on 1-simplices are considered. Extending this concept further allows for the inclusion of interactions within 2-simplices, encompassing every three qubits in the lattice.

\begin{figure}[h]
    \centering    \includegraphics[width=0.95\linewidth]{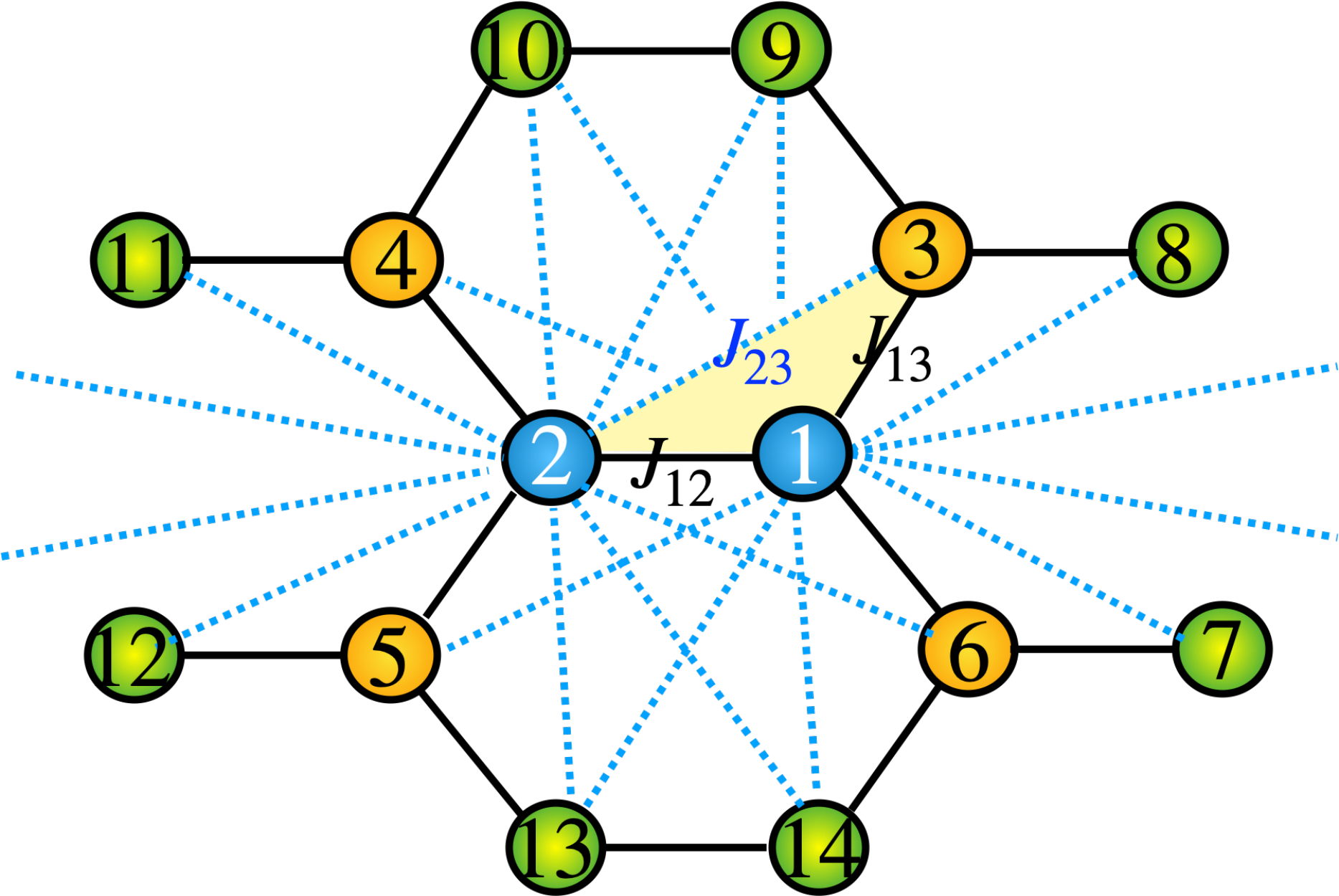}
    \caption{In a standard hexagonal lattice design, qubits are situated at the vertices, and couplers are placed at the midpoint of each solid line edge that connects the qubits. Dashed lines in this figure represent some unintentional couplings here; however, in principle, they exist everywhere in between every pair of qubits, making the lattice effectively an all-to-all lattice. A triangle highlighted in yellow exemplifies intentional couplings between qubit 1 and 2, as well as between 1 and 3. However, this triangle is closed by incorporating the unintentional interaction between qubit 2 and 3.} 
    \label{fig:hexagon}
\end{figure}

Figure \ref{fig:hexagon} illustrates a typical hexagonal lattice with qubits at the vertices, interacting along the edges with their nearest neighbors. Such configurations are instrumental in developing quantum error-correcting codes, like the triangular color code \cite{chamberland2020triangular}. Compared to traditional codes, this model boasts benefits in scalability and error threshold \cite{chamberland2020very,sahay2022decoder}.

Modeling a quantum processor begins with formulating Hamiltonian terms for every lumped electronic elements in the lattice and understanding the nature of their interactions. This Hamiltonian is then subject to various transformations and simplifications, ultimately being distilled into an effective Hamiltonian. In this reduced form, the couplers are typically omitted, and the focus is on the two-level qubits and their effective coupling strengths. This process establishes the computational lattice Hamiltonian.  The more comprehensive examination of the detailed circuit, which includes both qubits and couplers, will be reserved for the next section in which we study the example of a triangular lattice made with three qubits and three couplers located between every pair of qubits.

We investigate a qubit lattice as shown in Fig.~\ref{fig:hexagon}, characterized by its hexagonal patterns. The qubits are color-coded for clarity. Analyzing hexagonal geometries poses significant challenges due to their inherent complexity. To facilitate a comprehensive understanding, we initially focus on the simplest form of multi-qubit interaction: the three-qubit interaction. Our primary attention is on the blue qubits, labeled 1 and 2, surrounded by orange qubits numbered 3 to 6. These orange qubits exhibit \emph{intentional} coupling strengths with qubits 1 and 2, as indicated by solid lines. This lattice is deliberately designed and fabricated to establish intentional couplings between qubits. 

However, it is reasonable to consider that there is a level of weak coupling between every pair of qubits, possibly due to factors like weak capacitive interactions. The presence of such unintentional couplings between all pairs of qubits is significant. In Fig~\ref{fig:hexagon}, these \emph{unintentional} interactions are marked by dashed lines. The green qubits, labeled 7 to 14, while not having intentional couplings with qubits 1 and 2, are nonetheless interconnected through unintentional couplings. For instance, in a triangle highlighted in yellow, the coupling strengths $J_{12}$ and $J_{13}$ are intentional, whereas $J_{23}$ represents an unintentional but non-zero coupling. Theoretically estimating these coupling strengths necessitates a detailed understanding of the lattice's complete structure, including all couplers, which will be discussed in the next section.

The lattice, encompassing both intentional and unintentional interactions, forms a nearly all-to-all interaction network. A straightforward method to derive the lattice Hamiltonian starts with the Hamiltonian of electronic lumped elements in the full Hilbert space, typically limited to two-body interactions. The extensive Hilbert-space Hamiltonian (EHH) is described as~\cite{rasmussen21superconducting, krantz19a-quantum}:
\begin{align}\label{eq:Hj}
\frac{H}{\hbar}=&\sum_{q, n_q}\tilde{\omega}_q({n_q})|n_q\rangle\langle n_q| +\sum_{p,q, m_{p},n_q} \sqrt{(m_{p}+1)(n_{q}+1)}  \nonumber\\
&  \ \ \ \ \  \times J_{pq}^{m_{p}n_q} (|m_{p}, n_q+1\rangle\langle m_{p}+1,n_q|+h.c.),
\end{align}
with $p$ and $q$ are indices labeling qubits, while $m_p$ and $n_q$ denote their respective energy levels. The frequency $\tilde{\omega}_q({n_q})$ represents the effective energy difference between levels $n_q$ and $n_q+1$, i.e., $\hbar \tilde{\omega}_q({n_q})= E(n_q+1)-E(n_q)$. The tilde symbol differentiates effective qubit frequencies in the reduced circuit. The qubits exhibit a non-harmonic energy spectrum, leading to anharmonicity defined as $\tilde{\delta}_q = \tilde{\omega}_q(1) - \tilde{\omega}_q(0)$. 

The coupling strength, denoted as $J_{p\ \ q}^{m_p,n_q}$, represents the net coupling strength between qubits $p$ and $q$. This strength is influenced by both the energy difference between the qubits and the individual energy levels of each qubit. It comprises \emph{direct} and \emph{indirect} components, with the latter representing the coupling mediated by a coupler element, such as a transmission line, cavity, or another qubit. On the lattice, as described above, the solid edges represent intentional couplings and include the sum of both direct and indirect couplings. In contrast, the dashed lines, which indicate unintentional couplings, consist only of direct coupling.

 From the extensive Hilbert-space Hamiltonian of Eq.~\eqref{eq:Hj}, we can be projected onto the computational Hamiltonian. This projection usually maps the level repulsion/attraction between computational and non-computational levels into interactions involving more than two qubits in the Pauli basis. In a lattice consisting of a total of $N$ qubits, a block diagonalization transformation is employed to perform this projection \cite{nigg2012black-box,ansari2019superconducting}. The final Hamiltonian in the computational basis can be formally represented in the following effective Hamiltonian:
 
\begin{equation}
\label{eq:Hz}
\frac{H_{\rm eff}}{\hbar}= \sum_{\substack{\textup{qubit} \\ i}} \tilde{\omega}_i \frac{\hat{Z}_i}{2}  + \sum_{\substack{\textup{2-qubit} \\ i,j }} \alpha_{ij}\frac{\hat{Z}_i\hat{Z}_j}{4} + \sum_{\substack{\textup{3-qubit} \\ i,j,k }}  \alpha_{ijk}\frac{\hat{Z}_i\hat{Z}_j\hat{Z}_k}{8}
\end{equation}

In Eq.~\eqref{eq:Hz}, we limit our analysis to three-body entangling interactions. The numerous non-computational energy levels present in the original Hamiltonian~\eqref{eq:Hj} manifest in the effective Hamiltonian, Eq.~\eqref{eq:Hz}, as stray couplings, such as $ZZ$ and $ZZZ$. These  couplings do not cause state transitions in qubits, but they can accumulate phase errors in quantum states across different levels of entanglement. As these interactions are constantly active, their collective effect on the phase is continuous and pervasive.

In our study, we analyze the stray coupling $ZZ$ interaction between qubits $i$ and $j$ by separating the contributions of two-body from three-body interactions. The two-body contribution, denoted as $\zeta_{ij}$, is relatively straightforward to compute  ~\cite{{ku2020suppression},{xu2021zz-freedom},{roy2017implementation},{cai2021impact}}. 
The quantity $ \zeta_{ijk}$ is the three-body stray entanglement coupling (3-SEC) and quantifies the stray entanglement that can distort the intentional entanglement between three-qubits. In order to evaluate the influence of three-body interactions on the $ZZ$ interaction strength, it is essential to consider all potential third qubits, whether they interact intentionally or unintentionally with qubits $i$ and $j$. As depicted in Fig.~\ref{fig:hexagon}, we operate under the assumption that all qubits are interconnected. This leads us to infer that qubits $i$ and $j$ are part of triangular interactions with numerous third qubits, labeled as $k$ for clarity. Hence, the overall $ZZ$ interaction in a network of multiple qubits is calculated as
\begin{equation}
\alpha_{ij}=  \zeta_{ij}+\sum_k \zeta_{ijk}\label{eq:zz}
\end{equation}
 
The strength of the three-body operation  $\zeta_{ijk}$ will correspond to the following operator   $\hat{Z}_i \hat{Z}_j \hat{I}_k$ within the triangle formed by interacting qubits $i$, $j$, and $k$, which makes it eligible to be added to $ZZ$ interaction as three-body contribution.  

However $ZZ$ term is not the only stray coupling one can consider for a circuit of qubits. In the Hamiltonian~\eqref{eq:Hz} we generalized our analysis beyond the traditional focus of pairwise stray couplings to also include three-body stray entangling interactions, denoted as $ZZZ$. Essentially, in a lattice comprising $N$ qubits, effective Hamiltonian can encompass all stray entangling interactions, up to $N$-body interaction. Nonetheless, for practical simplicity, we examine  the effects up to three-body interactions within a lattice of multiple qubits.

\subsubsection{Two-body ZZ interaction}
In this section, we assess the parasitic interactions between the two blue qubits depicted in Fig.~\ref{fig:hexagon}. Initially, we conduct a  perturbative approach in estimating the coupling strength.

Within the perturbative regime of coupling strengths $|J_{pq}/[\omega_{p}-\omega_q]| \ll 1$, namely `dispersive regime',  one can show the following 2 useful symmetries:
\begin{align}\label{eq:J_relation}
J^{mn}_{ij} =J^{nm}_{ji}, &\ \ \forall n,m \in \{ 0,1,2,\cdots\} ,  \nonumber \\
J^{01}_{ij} + J^{10}_{ij}  = & J^{00}_{ij} + J^{11}_{ij}.
 \end{align}

In the following section, we introduce another relationship between different $J$ values, as detailed in Eq.~\eqref{eq.third}, after discussing the coupler specifics. In Appendix \ref{App.Identities}, we numerically validate the identities by simulating both sides of the equations and verifying their equalities.

By employing perturbation theory, one can calculate the two-body $ZZ$ coupling strengths up to the third order in $J$ as follows:
\begin{align}
 \zeta_{ij}/2 \equiv &  \frac{{(J_{ij}^{01}})^2}{\Delta_{ij}^{01}}-\frac{{(J_{ij}^{10}})^2}{\Delta_{ij}^{10}} \label{eq:zz2}\\
 \zeta_{ijk}/4 \equiv & \frac{J_{ij}^{00}J_{ik}^{00}J_{jk}^{00}}{\Delta_{ik}^{00}\Delta_{jk}^{00}}
+\frac{J_{ij}^{00}J_{ik}^{01}J_{jk}^{01}}{\Delta_{ik}^{01}\Delta_{jk}^{01}}\nonumber\\
&-\frac{J_{ij}^{01}J_{ik}^{00}J_{jk}^{10}}{\Delta_{ik}^{00}\Delta_{jk}^{10}}
-\frac{J_{ij}^{10}J_{ik}^{10}J_{jk}^{00}}{\Delta_{ik}^{10}\Delta_{jk}^{00}}-K{(1)}\label{eq:zz3}
\end{align}
with $\Delta_{ij}^{m_i n_j }= \tilde{\omega}_{i}(m_i)-\tilde{\omega}_j(n_j)$, defined as energy difference between two transitions $|m_i+1\rangle \leftrightarrow |m_i\rangle$ of  qubit $i$ and  $|n_j+1\rangle \leftrightarrow |n_j\rangle$ of qubit $j$. 

Using Eq.~\eqref{eq:J_relation} and  $\Delta_{ij}^{m_i n_j }= -\Delta_{ji}^{n_j m_i}$ one can show that $ \zeta_{ij}$ and $ \zeta_{ijk}$ are both symmetric under switching $i \leftrightarrow j$, (i.e. $\zeta_{ij}=\zeta_{ji}$ and $\zeta_{ijk}=\zeta_{jik}$), indicating that $ZZ$ interaction between qubits $i$ and $j$ is symmetric with respect to interchanging the qubit labels $i \leftrightarrow j$.

In Eq.~\eqref{eq:zz3} the quantity $K{(n)}$ is a scalar number and is defined as follows:
\begin{align}
    K{(n)} & =  - \sum_{\substack{i, j, k=\{1,2,3 \}\\(i<j;\  i,j\neq k)}} \frac{J_{ij}^{n n}J_{ik}^{n, 1-n}J_{jk}^{n, 1-n}}{\Delta_{ik}^{n, 1-n}\Delta_{jk}^{n, 1-n}}.\nonumber
\end{align} 
which turns out to depend in the leading order on $J^3/\Delta^2$. 

For example in a chain of qubits, without considering unintentional couplings, the three-body and higher are identically zero, which is due to the absence of symplectic interaction. One can see in Eq.~\eqref{eq:zz3} that even a small triangularity makes $ \zeta_{ijk}$ nonzero, which can even be significant for nearly in resonance qubits. 

\subsubsection{Three-body ZZZ interaction}
In a circuit with at least three qubits in triangular interactions,  perturbation theory  estimates the following $ZZZ$ interaction strength which is a one-shot three-body interaction:  
\begin{eqnarray}\label{eq:zzz}
\alpha_{ZZZ} /8 &= &  \sum_{n=0,1}K{(n)} 
\end{eqnarray} 

The perturbative three-body $ZZZ$ interaction is generally considered to be weaker than two-body $ZZ$ interactions. This is attributed to the higher-order dependence of the three-body interaction compared to the two-body interaction within the perturbative parameter $J/\Delta<1$. Such an observation is consistent with many-body localization (MBL) theory, as highlighted in various key studies. MBL theory suggests a distinct hierarchy of interaction scales in weakly interacting spins. In a triangular configuration of superconducting qubits, the nearest-neighbor $ZZ$ interactions are typically the most pronounced, with the $ZZZ$ couplings being considerably weaker.

However, beyond the dispersive regime, qubit-qubit interactions is stronger, as detailed in \cite{ansari2019superconducting}. For instance, when the frequency of the coupler approaches that of the qubits, the strength of both two-body and three-body interactions may challenge the established MBL hierarchy. Specifically, this could lead to a scenario where $\alpha_{ZZZ}$ is equal to or greater than $\alpha_{ZZ}$, a phenomenon we refer to as $ZZZ$ \emph{superiority}. Interestingly, this superiority might even occur within the dispersive regime, as shown in Fig.~\ref{fig:zzzsuper}.

\begin{figure}[h]
    \centering
     \caption{Phase diagram of $ZZZ$ superiority where $|ZZZ|>|ZZ|_{\rm max}$ as a function of coupling strength $|J/\Delta|$ and anharmonicity $|\delta/\Delta|$ at different  qubit frequency detunings with $\Delta=\Delta_{32}$. The shaded area enclosed represents the region (red using perturbative SWT and green from non-perturbative numerical analysis) where three-body $ZZZ$ interaction is superior to two-body $ZZ$ interaction.}
    \label{fig:zzzsuper}
\end{figure}

To illustrate this, we consider a triangular circuit comprising three qubits, each with a uniform pairwise coupling strength $J$ and anharmonicity $\delta$. By comparing the strength of the $ZZZ$ interaction as described in Eq.~(\ref{eq:zzz}) with the $ZZ$ interaction as per Eq.~(\ref{eq:zz}), we can identify the conditions under which $ZZZ$ superiority emerges. Figure \ref{fig:zzzsuper} (a,b,c,d) display these conditions for various frequency settings of $\Delta_{31}$, including $-\Delta_{32}$, $0.5 \Delta_{32}$, $1.2 \Delta_{32}$, and $3 \Delta_{32}$. 
These scenarios are represented analytically in red regions using Schrieffer-Wolff Transformation (SWT) and are compared with green regions that depict numerical analysis results by diagonalizing the Hamiltonian Eq.~\eqref{eq:Hj}.
The discrepancy arises from the breakdown of the perturbative SW transformation beyond dispersive regime.

\section{A three-qubit lattice}\label{sec:two1}

\begin{figure}[h!]
    \centering
    \includegraphics[width=0.9\linewidth]{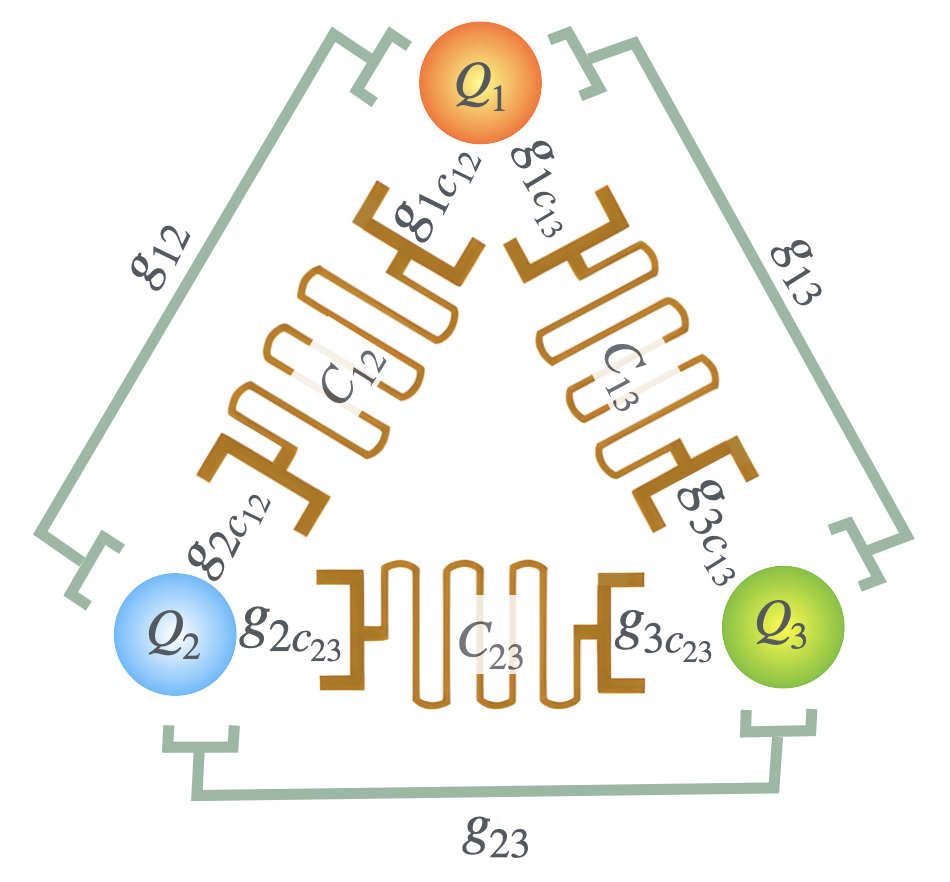}
    \vspace{-0.1in}
    \caption{A typical triangle of three qubit $Q1, Q2$, and $Q3$, which pairwise interact in a combination of direct couplings $g_{ij}$ and indirect couplings $g_{ic_r}$ via the couplers $C_r$. }
    \label{fig:3qubit}
\end{figure}

Considering that an all-to-all lattice of qubits can be divided into numerous triangular building blocks, this section will specifically focus on one such triangle. In this triangle, three qubits are pairwise connected through three couplers. Moreover, we also account for direct couplings between the qubits, as depicted in Fig.~\ref{fig:3qubit}. The effective Hamiltonian is calculated using both perturbative and non-perturbative approaches. A systematic comparison is then made between scenarios with and without the inclusion of a third qubit. This comparative analysis is instrumental in understanding the nuances of qubit interactions within such a lattice configuration. 

Under the parameters typically used in quantum computing we interestingly uncover the alteration of multi-qubit interaction strength hierarchy. This provides key insights into the complex dynamics of qubit interactions and challenges prevailing assumptions about how these interactions are organized and prioritized within quantum systems.

In Fig.~\ref{fig:3qubit}, qubits are represented by circles and are labeled as $Q_q$,  $q=\{1,2,3\}$. The rectangles symbolize the couplers $C_r$, with $r=\{{12},{23},{13}\}$ indicating the labels of the two qubits they mutually couple to intentionally. The coupling strength between a qubit $Q_q$ and a coupler $C_r$ is denoted as $g_{qc_r}$. Qubits interact indirectly through their shared coupler, yet they can also have direct interactions, such as capacitive ones. The direct coupling between qubit $q$ and another qubit $q'$ is represented as $g_{qq'}$. 

For simplicity in our analysis, we assume all couplers function like harmonic oscillators, akin to transmission lines or cavity modes. In the model for Fig.~\ref{fig:3qubit}, we neglect the weak interactions between qubits and distant resonators. This means we consider $g_{1c_{23}} = g_{2c_{13}} = g_{3c_{12}}$ as negligible or equal to zero. Thus, we express the circuit Hamiltonian in our study as follows:
\begin{align}
\label{eq:triangle}
\frac{H}{\hbar}&= \sum_{q} \omega_q \hat{a}_q^{\dagger}\hat{a}_q+\frac{\delta_q}{2}\hat{a}_q^{\dagger}\hat{a}_q(\hat{a}_q^{\dagger}\hat{a}_q-1) + \sum_{r}\omega_{c_r}\hat{c}_r^{\dagger}\hat{c}_r \nonumber \\
&+\sum_{r, q,q'} \left[  g_{qq'}(\hat{a}_{q}-\hat{a}_{q}^{\dagger}) + g_{qc_r}(\hat{c}_r-\hat{c}_r^{\dagger}) \right] (\hat{a}_{q'}-\hat{a}_{q'}^{\dagger}) 
\end{align}
with $\omega_{q}$ being qubit bare frequency, $\delta_q$  qubit anharmonicity, and $\omega_{c_r}$ coupler frequency.  In order to obtain the analytical Hamiltonian between qubits in the computational subset, first we restrict circuit parameters within the dispersive regime i.e. $g_{qc_r}\ll |\omega_{c_r}-\omega_q|$, and using the Schrieffer-Wolff transformation, which helps to make the Hilbert space of harmonic couplers separable from qubit subset, therefore couplers can be safely eliminated and their effect renormalizes bare qubit frequency into their dressed values:
 \begin{equation}
\tilde{\omega}(n_q)=\left(\omega_q+ \frac{(n_q-1) \delta_q}{2} \right) n_q  -\sum_r \frac{g^2_{qc_r} n_q}{\Delta_{c_r q}(n_q-1) }   ,
\end{equation}
with $r$ summing over  those resonators that  interact with  qubit $q$ via $g_{qc_r}$.  We define $\Delta_{c_{r} i}(m_i)\equiv\omega_{c_r}-\omega_i-m_i\delta_i$. 

Two qubits $q$ and $q'$ that share the same coupler $C_r$ and couple to it by the strengths $g_{qc_r}$ and $g_{q'c_r}$, effectively interact with one another by the following effective strength:
\begin{equation} \label{eq. Jeff}
J_{qq'}^{m_q n_{q'}}=g_{qq'}-\frac{g_{qc_r}g_{q'c_r}}{2}\left(\frac{1}{\Delta_{c_rq}(m_q)}+\frac{1}{\Delta_{c_rq'}(n_{q'})}\right),
\end{equation}
with $g_{qq'}$ being direct (i.e. capacitive) coupling strength between the two qubits and the second term being the perturbative indirect coupling strength via the shared coupler.

A few lines of algebra proves the validity of the two identities in Eq.~\eqref{eq:J_relation}, and a third one below, which turn out to be useful set of relations for simplifying problems. 
\begin{align}\label{eq.third}
J^{01}_{ij} - \beta_{ij} J^{10}_{ij}  = & \left( 1-\beta_{ij}\right) J^{00}_{ij}.
 \end{align}
with $\beta_{ij} = \frac{\delta_{j}}{\delta_{i}} \frac{\Delta_{ci}(0)}{\Delta_{cj}(0)} \frac{\Delta_{ci}(1)}{\Delta_{cj}(1)}$ and  $\beta_{ji}=1/\beta_{ij}$.

\begin{figure}[ht] 
\includegraphics[width=0.24\textwidth]{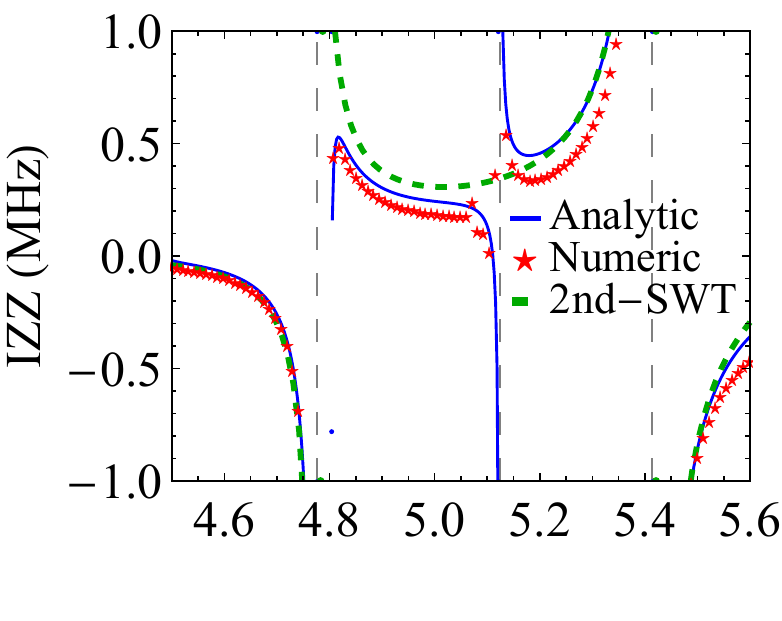}\put(-123,90){\textbf {(a)}}
\includegraphics[width=0.24\textwidth]{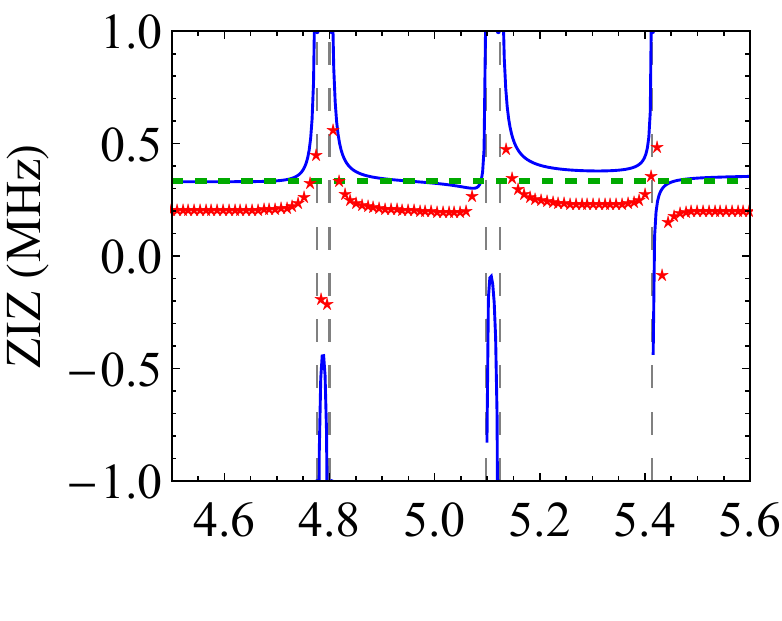}\put(-123,90){\textbf {(b)}}\\
\vspace{-0.1in}
\includegraphics[width=0.24\textwidth]{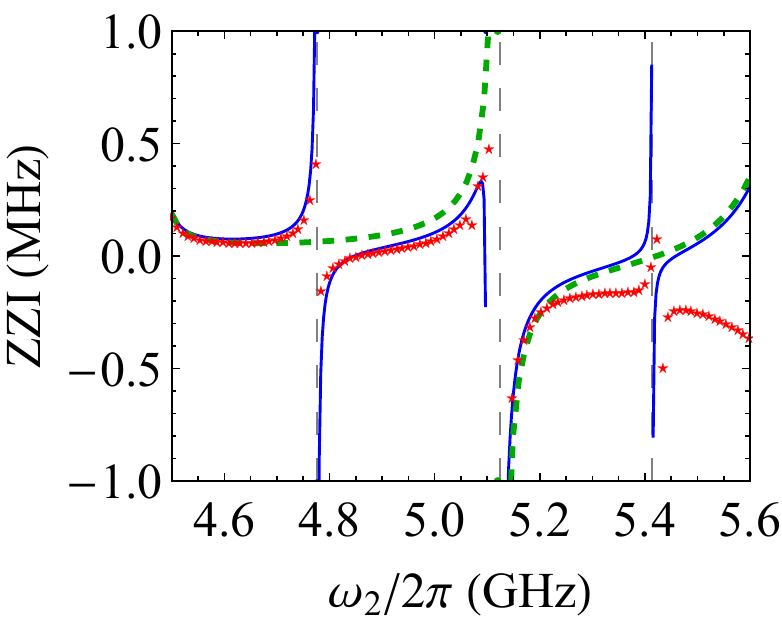}\put(-123,90){\textbf {(c)}}
\includegraphics[width=0.24\textwidth]{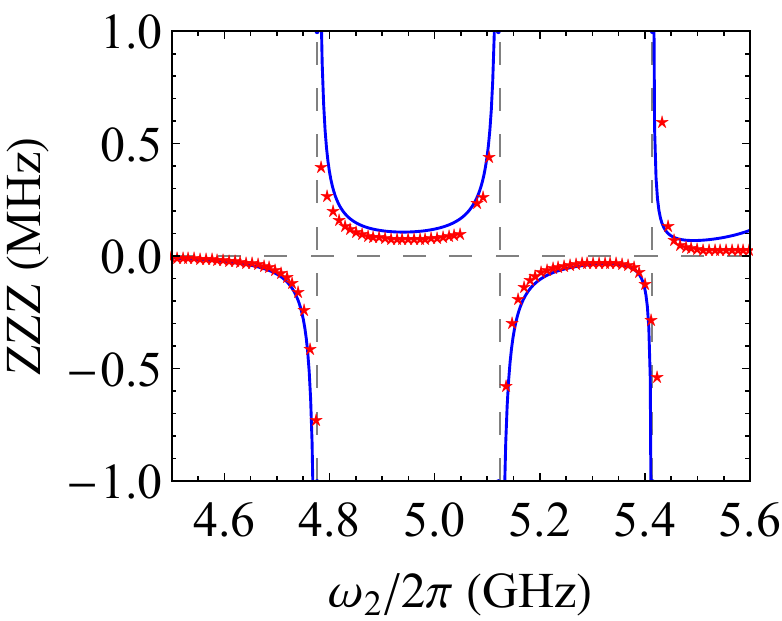}\put(-123,90){\textbf {(d)}}
\vspace{-0.1in}
\caption{Two-body $ZZ$ and three-body $ZZZ$ in a three qubit circuit with all-to-all  connectivity. Solid lines represent the analytical results for $ZZ$ in Eq.~\eqref{eq:zz} and $ZZZ$ in Eq.~\eqref{eq:zzz}, while the dashed lines correspond to the second-order SWT results in Eq.~\eqref{eq:zz2} for $ZZ$, with no contribution to $ZZZ$ since both first and second-order terms are zero, while asterisks show numerical simulation by diagonalizing Eq.~\eqref{eq:triangle}. Circuit parameters are listed as follows: $\omega_1/2\pi=4.8$ GHz, $\omega_3/2\pi=5.1$ GHz, $\delta_1/2\pi=\delta_2/2\pi=\delta_3/2\pi=-330$ MHz, $\omega_{c_{12}}/2\pi=5.8$ GHz, $\omega_{c_{23}}/2\pi=6.1$ GHz, $\omega_{c_{13}}/2\pi=5.7$ GHz, $g_{1c_{12}}/2\pi=g_{1c_{13}}/2\pi=g_{2c_{12}}/2\pi=g_{3c_{13}}/2\pi=85$ MHz, $g_{2c_{23}}/2\pi=g_{3c_{23}}/2\pi=102$ MHz, direct qubit-qubit coupling $g_{12}/2\pi=g_{23}/2\pi=4$ MHz and $g_{13}/2\pi=6$ MHz.\label{fig:zzzvszz}}
\end{figure}

We can assess both two-body and three-body interactions within the simplified three-qubit circuit, utilizing parameters determined in the dispersive regime. Figures~\ref{fig:zzzvszz} (a-c) illustrate the dependency of the $ZZ$ interaction on the frequency (whether bare or dressed) of a specific qubit, in this case, $Q_2$. Figure~\ref{fig:zzzvszz}(d) presents the results of the three-body $ZZZ$ interaction as per Eq.~\eqref{eq:zzz}. The precision of our theoretical derivation is corroborated by the results of our numerical simulations.

Figures~\ref{fig:zzzvszz} (a-c) display the theoretical estimation of two-qubit $ZZ$ interactions among different pairs in a triangular qubit configuration, with the variation of frequency in qubit $Q2$. The green dashed lines represent second-order perturbative results, which take into account only pairwise interactions only up to $J^2$ \cite{magesan2020effective}. The blue solid lines show perturbative results based on Eqs.~(\ref{eq:zz2},\ref{eq:zz3}) up to third order in $J$ in which naturally three-qubit correction terms $K(n)$ are included. Furthermore, red asterisks indicate our numerical non-perturbative results, obtained using the {\href{cirqubit.com} {CirQubit}} software. This comprehensive approach provides a deeper understanding of the interaction dynamics within the qubit triangle.

Comparing the two perturbative analyses — one focusing solely on pairwise interactions and the other encompassing three-body interactions — provides valuable insights into the influence of multi-qubit interactions on quantum device performance. In scenarios where the frequency of $\omega_2$ is considerably distinct from the other two qubits, the analytical and numerical results align closely. This suggests that the impact of three-body corrections is relatively minor in such domain of frequencies.

However, as the frequency of $Q2$ approaches that of another qubit, the results based solely on pairwise interactions deviate from the other two results, the three-body formulation and numerical data. Interestingly, our equations (\ref{eq:zz2},\ref{eq:zz3}), which incorporate three-body interactions, continue to provide a reasonable approximation to the numerical interaction strength, even near points of discontinuity. The numerical Hamiltonian evaluates finite points in proximity to divergences, typically where quantum states are nearly degenerate.

Interestingly, as shown in Fig.~\ref{fig:zzzvszz}(a) and (c), the interactions $IZZ$ and $ZZI$ can be nullified at two distinct $Q2$ frequencies. These specific frequencies are accurately predicted by Eqs. (\ref{eq:zz2},\ref{eq:zz3}), underscoring the effectiveness of these equations in capturing the nuances of multi-qubit interactions in quantum systems.

In Fig.~\ref{fig:zzzvszz}(b), the $ZIZ$ interaction predominantly maintains a positive value across a wide range of $Q2$ frequencies, with the exception of a few specific frequencies where the system is nearly degenerate. It's important to note that the primary contribution to the $ZIZ$ interaction originates from $Q1$ and $Q3$, rather than $Q2$. The  gap of approximately 100 kHz between the analytical formula and the numerical simulation results,  persists even in a well-defined dispersive regime where $Q2$'s frequency is distinctly separate from the other qubits. This discrepancy can be attributed to the omission of higher-order corrections in the Schrieffer-Wolff transformation. This transformation is used perturbatively to decouple resonators. More details can be found in Appendix~\ref{app:ZIZD}. By including higher-order decoupling of resonators, the results become more consistent with the numerical simulations, thereby enhancing the accuracy of the analytical approach.

Our non-perturbative analysis, illustrated by the red asterisks in Fig.~\ref{fig:zzzvszz}(d), supports the conventional hierarchy where the strength of $ZZZ$ interactions is weaker than that of $ZZ$ interactions in circuits with specific frequency arrangements, such as $(\omega_2<\omega_1<\omega_3$ or $\omega_1<\omega_3<\omega_2)$.  However, under certain frequency configurations, notably $(\omega_1<\omega_2<\omega_3)$, the strength of $ZZZ$ interactions can exceed that of $ZZ$ interactions like $ZZI$.

Our findings in examining lattice stray couplings challenge the widely held belief that higher-order interactions are always much weaker. This suggests that there might be an oversight in conventional quantum simulations. Traditionally, efforts have been focused primarily on minimizing $ZZ$ interactions as a means to enhance gate fidelity. However, our study indicates that this approach could be misleading. In certain circuit configurations, $ZZZ$ interactions can play a significant role, potentially leading to unwanted entanglement. This realization underscores the need for a more comprehensive consideration of various interaction types in the design and analysis of quantum systems, emphasizing that a broader perspective is essential for optimizing the performance and accuracy of quantum computing technologies.

Briefly, our study delves into the expanded scope of stray couplings beyond the $ZZ$ interaction, focusing specifically on the three-body $ZZZ$ interaction in an existing multi-qubit device. This approach marks a departure from previous research, which predominantly concentrated on 2-body interactions among three qubits and introduced non-simultaneous 3-qubit gates. Our investigation thus broadens the understanding of multi-qubit interactions, particularly emphasizing the significance and characteristics of three-body interactions in quantum computing systems.

\section{$ZZ$ gate in a triangle}\label{sec:two2}

In quantum circuits, the operation of a two-qubit gate, such as one involving qubits 2 and 3 in a triangular lattice depicted in Fig.~\ref{fig:3qubit}, can be significantly affected by the presence of additional qubits, like qubit 1 in this scenario. The influence of such external qubits on gate performance typically arises through two main categories of interactions: (1) two-body interactions, which occur between 1 and 2, or 1 and 3, and (2) three-body interactions, involving all three qubits 2, 3, and 1 simultaneously.

Contrary to common belief, which regards three-body interactions as weaker and often negligible compared to two-body interactions, this section emphasizes the need to consider three-body interactions in circuits. This consideration helps enhance two-qubit gate fidelity, furthermore it may pave the road toward introducing meaningful instantaneous three-qubit gates.  It's useful to categorize decoupling strategies into two distinct types: (1) hard decoupling  --this method aims to find an optimal operating point where the qubits involved in the gate operation are completely isolated, preventing even weak interactions with ungated qubits. The goal is to reach a state in which the coupling strength between the gated qubits and any additional, non-participating qubit is effectively null. (2) Soft decoupling  -- Rather than striving for strict decoupling, we may focus on reducing stray couplings.

A critical aspect of our research involves conducting an extensive analysis of three-body lattice interactions among the qubits and assessing their impact on the performance of two-qubit gates.  Each strategy presents its own set of advantages and challenges. The choice between hard and soft decoupling often depends on the specific requirements of the quantum circuit and the nature of the qubits involved.  We will understand in this section that the concept of hard decoupling  might not always be feasible or even advantageous. Thus, our aim will be to concentrate on mitigating the effects of these interactions instead of trying to eliminate them completely. This method is particularly useful in complex quantum circuits where strict decoupling  is challenging to achieve, or where selective interactions can contribute positively to the system's functionality.

\subsection{Hard $J$-decoupling}\label{sec. strict}

Consider the three-qubit system shown in Fig.~\ref{fig:3qubit}. Each qubit in this system possesses multiple energy levels. Within the triangular lattice structure, a qubit $Q_i$ can be excited from an energy level $n_i$ to $n_i+1$. This excitation can occur concurrently with the de-excitation of another qubit $Q_j$ from the energy level $m_j+1$ down to $m_j$. The strength of coupling $J_{i\  j}^{m_i, n_j}$ depends on a range of parameters related to both  qubits and couplers.

\begin{figure}[t]
\centering
\includegraphics[width=0.492\linewidth]{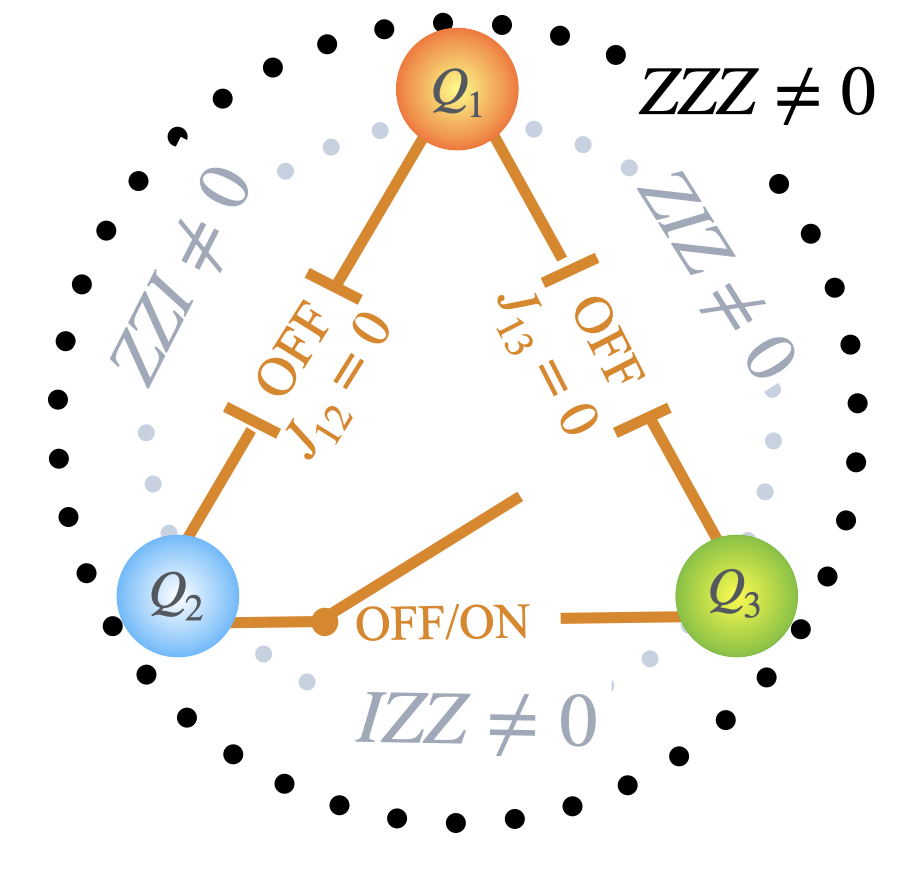} \includegraphics[width=0.492\linewidth]{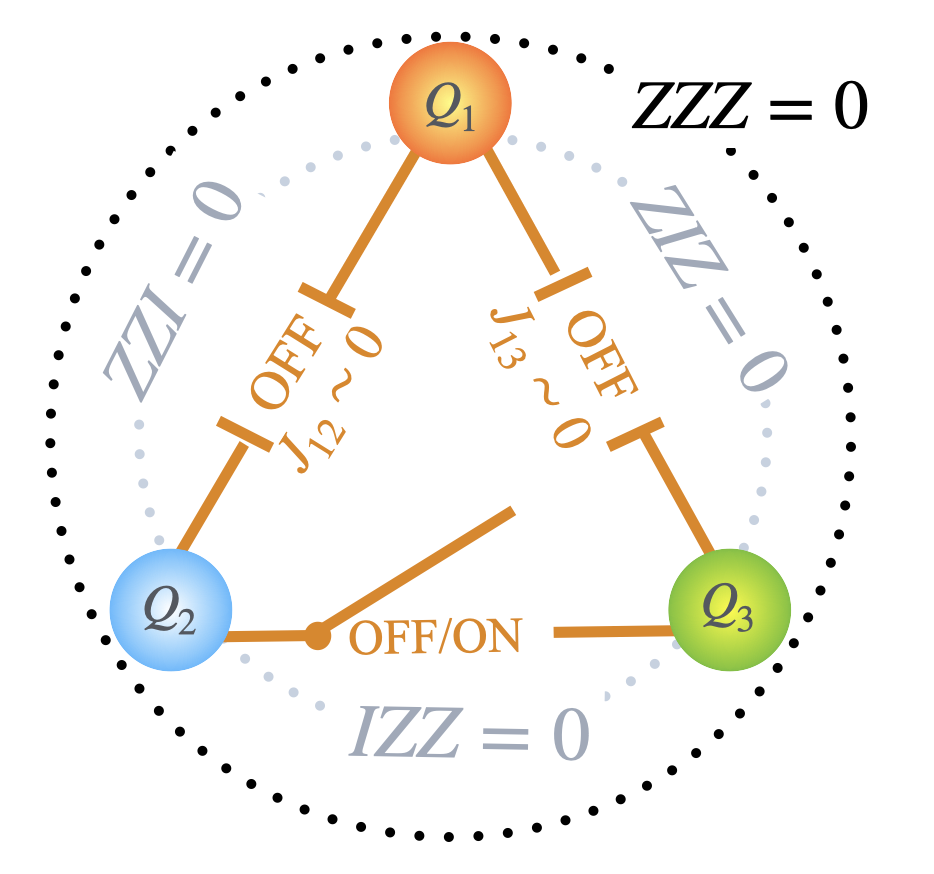}
\put(-220,0){\textbf {(a)}}
\put(-100,0){\textbf {(b)}}
\vspace{-0.1in}
\caption{(a) Hard $J$-decoupling at $J_{12}=J_{13}\approx0$, (b) Soft $J$-decoupling at $J_{12}\sim$ 1 MHz and $J_{13}=0$}
\label{fig:hardsoft}
\end{figure}

\begin{figure*}
\centering
\includegraphics[width=1\linewidth]{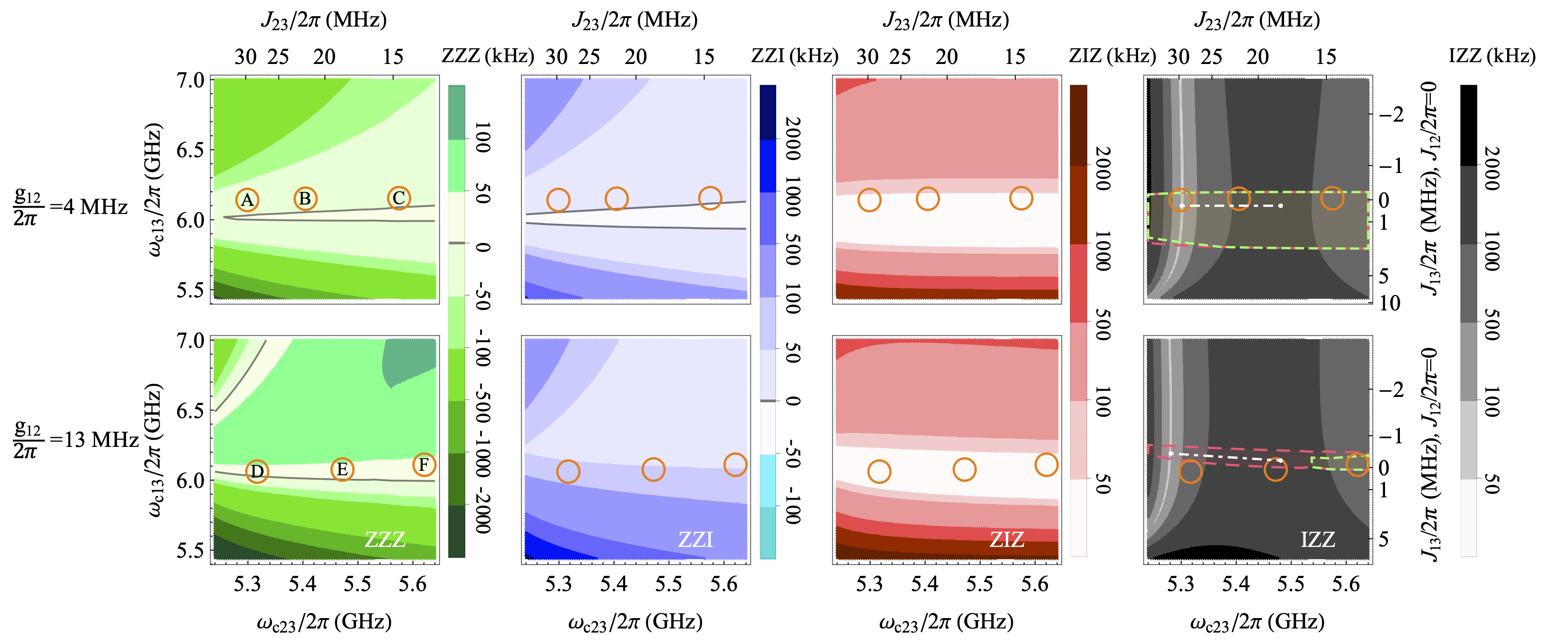}
\vspace{-0.1in}
\caption{Stray couplings in the triangular circuit with all-to-all connectivity calculated by numerically diagonalizing the Hamiltonian Eq.~\eqref{eq:triangle}, as a function of the frequencies of the resonators C23 and C13 i.e. $\omega_{c2}$ and $\omega_{c3}$ respectively. The first row is at  $g_{12}/2\pi=4$ MHz with nearly zero $J_{12}$ and $J_{13}$ at spot A with coupler frequencies $(\omega_{c12},\omega_{c23},\omega_{c13})/2\pi=(6.66,5.299,6.141)$ GHz,  B: $(6.67,5.405,6.15)$ GHz and C:  $(6.68,5.275,6.154)$ GHz. The second row is at $g_{12}/2\pi=13$ MHz with nearly zero $J_{12}$ and $J_{13}$ at spot D: $(5.438,5.317,6.021)$ GHz, E: $(5.443,5.472,6.076)$ GHz and F: $(5.446,5.621,6.098)$ GHz. The shaded regions denote two intersections: $|ZZI|\displaystyle \cap |ZIZ|<50$ kHz (light red) and ${\rm max}(|ZZI|,|ZIZ|,|ZZZ|)<50$ kHz (light green), both on top of the $IZZ$ interaction. Other circuit parameters are the same as in Fig.~\ref{fig:zzzvszz} with $\omega_2/2\pi=5.0$ GHz.}
\label{fig:ffn_z2d}
\end{figure*}

The common belief is that the optimum operating point for a two-qubit gate within a triangular circuit is where two couplings are effectively turned off (`OFF'), while the third remains active (`ON'), thus establishing an `OFF-OFF-ON' state. To analyze the operating point of this 2-qubit gate in presence of a third interacting qubit, first we begin by setting $J_{12}^{00}$ to zero, achievable at certain frequencies of the coupler $C_{12}$. This is what is namely `hard $J$-decoupling', which is diagrammatically sketched in Fig.~\ref{fig:hardsoft}(a). In our simulation we apply the condition step by step, first we turn off coupling between Q1 and Q2, then between Q1 and Q3, and finally switch on and off $CZ$ gate between Q2 and Q3. Figure~\ref{fig:ffn_z2d} showcases circuits with parameters outlined in the caption, all configured such that $J_{12}^{00}$ is nullified. We then focus on turning off the second coupling, say $J_{13}^{00}$, with its values indicated on the right axes of the figure. Simultaneously, the third interaction, $J_{23}^{00}$, marked on the upper axes, is selected to be strong. The variation of these couplings is achieved by adjusting the frequencies of couplers $C_{23}$ and $C_{13}$, as shown on the lower and left axes respectively.

The color bars on the figure represent the strengths of two- and three-body stray couplings. Moving from left to right, the column corresponds to the following interactions: $ZZZ$ (in green), $ZZI$ (in blue), $ZIZ$ (in red), and $IZZ$ (in grey), which we evaluate non-perturbatively, \cite{cederbaum1989block,magesan2020effective,xu2021zz-freedom,cirqubit}. Figure~\ref{fig:ffn_z2d} is a valuable tool for identifying the optimal settings for achieving the desired gate operations, particularly where multiple interactions coexist and influence each other. This level of detailed analysis is crucial for the precise tuning of quantum circuits, enhancing their performance and reliability in various quantum computing applications.

We evaluate the interaction strengths of the $ZZ$ and $ZZZ$ for the condition of $J_{12}^{00}=0$ , ensuring $Q1$ and $Q2$ do not interact. In Fig.~\ref{fig:ffn_z2d} we achieve the decoupling  by testing two different values for direct capacitive coupling between $Q1$ and $Q2$, i.e. $g_{12}/{2\pi}=4,13$ MHz in the upper and lower rows, respectively. Given the circuit parameters selected, the next coupling parameter to be neutralized is $J_{13}^{00}=0$. The goal of this setting is to either preserve or enhance the coupling between $Q2$ and $Q3$, represented by $J_{23}^{00}$. This is achieved by tuning the frequency of the $C_{13}$ coupler within a range of 5.4~GHz to 7~GHz. As a result, the coupling strength $J_{13}/2\pi$ varies from $-3$~MHz to $+9$~MHz, reaching zero at the midpoint. Consequently, any point along the $J_{13}=0$ line is characterized by a precise OFF-OFF interaction, which is critical for the intended gate operation. 

Regarding the exclusive ON coupling, $J_{23}^{00}$, we modify the frequency of the $C_{23}$ coupler by approximately 400 MHz on the lower axis. This adjustment leads to strong $J_{23}^{00}/2\pi$ values that range from 15~MHz to 30~MHz. We have marked three sample points with red circles and labeled them A to C in the upper row corresponding to the weaker $g_{12}$, D to E in the lower row corresponding to the stronger $g_{12}$.

In exploring the stray $ZZ$ and $ZZZ$ couplings, we focus on the horizontal line corresponding to $J_{13}=0$ in Fig.~\ref{fig:ffn_z2d}. This line signifies OFF-OFF interaction points achieved by setting both $J_{12}^{00}$ and $J_{13}^{00}$ to zero. While this condition is intended to decouple $Q1$ from $Q2$ and $Q3$, the figure illustrates that residual stray couplings still exist between $Q1$ and the other two qubits, notably in $ZZZ$ and $ZZI$ configurations. This observation is crucial as it demonstrates that hard decoupling  of seemingly non-interacting qubits does not necessarily lead to the elimination of stray couplings.

The persistence of these stray couplings is attributed to a significant factor. Although decoupling of two qubits, labeled $i$ and $j$, is achieved by setting their $J_{ij}^{00}$ (computational level coupling) to zero, these qubits remain indirectly coupled at their non-computational levels. These levels include couplings such as $J_{ij}^{01}$, $J_{ij}^{10}$, and others. It is these higher-order couplings that the stray interactions depend upon. Thus, even when qubits are seemingly decoupled at their fundamental computational states, indirect couplings at other energy levels persist, leading to stray interactions that must be considered in quantum circuit design and operation. Details can be found in Appendix~\ref{app:Jdis}.

The central aim of our study is to pinpoint `sweet spots' where both two-body and three-body stray couplings are minimal, specifically below 50~kHz. This quest is particularly evident in the last column of Fig.~\ref{fig:ffn_z2d}. Here, we highlight regions with low two-body $ZZ$ interactions, where the maximum of $|ZZI|$ and $|ZIZ|$ remains under 50~kHz, as indicated by red dashed boxes. Furthermore, areas showing both minimal $ZZZ$ and residual $ZZ$ couplings, characterized by the condition ${\rm max}\{|ZZI|,|ZIZ|,|ZZZ|\}<50$~kHz, are enclosed in green outlines.

The upper and lower plots in this column represent scenarios of weak and strong capacitive coupling between $Q1$ and $Q2$, denoted as $g_{12}$. In the case of weak $g_{12}$, all red dashed boundaries converge, forming an extensive frequency range conducive to effectively decoupling $Q1$ from the $Q2$-$Q3$ pair without significant stray couplings. This finding is particularly relevant for quantum devices designed for precise two-qubit gate operations, like the $CZ$ gate. Devices with such characteristics can smoothly transition from low to high $ZZ$ interaction values while keeping spectator errors — errors affecting qubits not directly involved in the gate operation — to a minimum. This aspect is crucial for improving the overall performance and reliability of quantum gates in computational tasks.

In Fig.~\ref{fig:ffn_z2d}, the white dotted-dashed line within the red boundary, which represents a zone of low $ZZ$ interaction, portrays an ideal trajectory for a parasitic-free $ZZ$ gate operation between $Q2$ and $Q3$. This path transits from nearly zero on the left side to a strong value on the right side. Crucially, along this line, $J_{12}^{00}$ and $J_{13}^{00}$ are effectively zero, and both $ZZI$ and $ZIZ$ remain below 50~kHz, suggesting an efficient decoupling of $Q1$ from the actively gated qubits, $Q2$ and $Q3$. 

However, it's critical to take a careful look at the impact of $ZZZ$ interactions for the gate. A comparison between the upper and lower plots in the last row of Fig.~\ref{fig:ffn_z2d} reveals critical differences between weak and strong $g_{12}$ regimes. In the weak $g_{12}$ scenario (upper plot), the ideal gate trajectory is entirely within the green boundary, indicating that the three-body interaction is maintained below 50~kHz. This implies a safer operating zone where both two-body and three-body parasitic interactions are minimized. In contrast, in the strong direct coupling between $Q1$ and $Q2$ (lower plot), the gate path falls outside this low $ZZZ$ interaction area. Here, the $ZZZ$ value can potentially exceed 100~kHz, suggesting a heightened vulnerability to parasitic three-body interactions.

This observation reveals that assuming three-body $ZZZ$ interactions are weaker than two-body $ZZ$ interactions can lead to significant oversights in gate design. Suppressing parasitic two-body $ZZ$ interactions may inadvertently introduce substantial errors from strong three-body interactions among presumed noninteracting qubits. Therefore, accurately assessing three-body interactions is essential to avoid flawed designs and achieve high-fidelity quantum operations.

\begin{figure*}[t]
\centering
\includegraphics[width=1\linewidth]{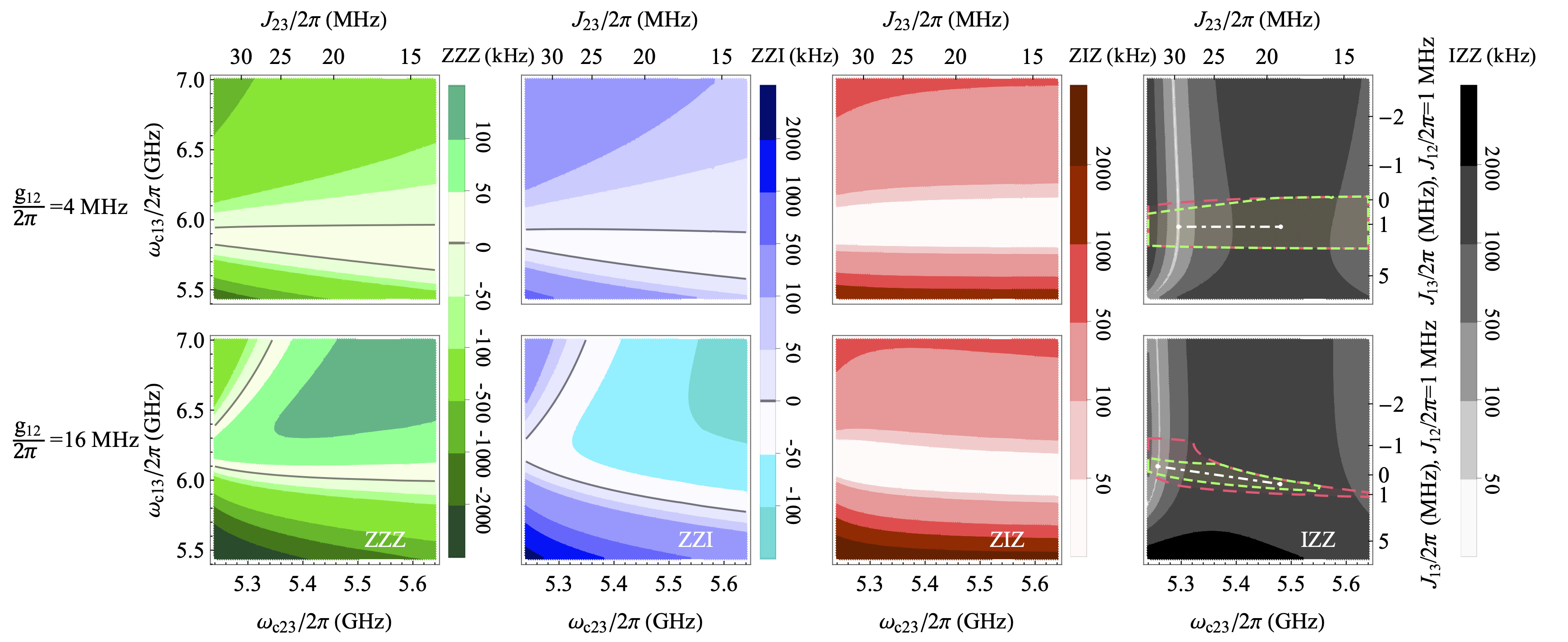}
 \caption{Stray couplings in the triangular circuit with all-to-all connectivity calculated by numerically diagonalizing the Hamiltonian Eq.~\eqref{eq:triangle},  as a function of the frequencies of the resonators C23 and C13 $\omega_{c23}$ and $\omega_{c13}$ respectively. The first row is at  $g_{12}/2\pi=4$ MHz with weak $J_{12}/2\pi\approx1$ MHz at $\omega_{c12}/2\pi=6.2$ GHz. The second row is at $g_{12}/2\pi=16$ MHz with weak $J_{12}/2\pi\approx1$ MHz at $\omega_{c12}/2\pi=5.3$ GHz. Other circuit parameters are the same as Fig.~\ref{fig:ffn_z2d}.}
    \label{fig:fnn_z3d}
\end{figure*}

\subsection{Soft $J$-decoupling}

The common desire in multi-qubit circuit design is that during the application of a two-qubit gate, other qubits are silenced. This is usually interpreted as turning off $J$ interaction between the gated qubits and other qubits.  In the previous section we examine this concept in the context of a triangular lattice by setting two pairwise interactions to zero and amplifying the third. In the previous section we referred to this concept as hard $J$-decoupling. However, our  detailed analysis reveals that the situation is significantly more complex than nullifying $J$ interaction.

In this section, we explore the same OFF-OFF-ON interaction between $Q2$ and $Q3$ in an alternative setup: Rather than enforcing a complete absence of interaction of these two qubits with $Q1$, we allow some weak $J$-interactions to take place if they contribute to reducing parasitic interactions. Our goal shifts towards minimizing stray coupling strengths to the extent that they either disappear or become negligible (i.e., less than 50~kHz). Enforcing zero interaction between qubits is limited in scope; it only prevents interactions between two specific energy levels of one qubit with another, which is inadequate to prohibit transitions across all energy levels. For instance, while hard decoupling might set $J_{12}^{00}$ to zero, other coupling strengths like $J_{12}^{01}$ could still be non-zero. The so-called `soft $J$-decoupling', is sketched diagrammatically in Fig.~\ref{fig:hardsoft}(b).

This realization implies that the idea of completely nullifying $J$-interaction between qubits might be overly simplistic, if not altogether unachievable. Recognizing this limitation can shift the focus towards managing stray interactions rather than striving for complete non-interaction. In other words, we focus  on the impact of relaxing one of the two OFF couplings to a near-OFF state, denoted as $\widetilde{\textup{OFF}}$. This softer approach to qubit decoupling, or `soft $J$-decoupling', is explored as a potentially more effective and realistic strategy for enhancing $ZZ$ interaction between $Q2$ and $Q3$, as opposed to the hard decoupling strategy previously considered.

\emph{Soft $J$-decoupling Criterion}: By setting the interaction between $Q1$ and $Q2$ to $J_{12}^{00}/2\pi=1$ MHz, rather than completely zeroing it, we introduce a `soft decoupling ' criterion. Figure~\ref{fig:fnn_z3d} illustrates how this  can  impact both two and three-body stray couplings in different circuit configurations: First (Second) row denotes a circuit with a weak (strong) direct capacitive coupling between $Q1$ and $Q2$, $g_{12}=4 (16)$~MHz. Such strength occurs at $\omega_{C_{12}}/2\pi=6.2 (5.3)$~GHz. 

The identification of `safe zones' for low stray couplings is crucial, which is identifiable in the last column of Fig.~\ref{fig:fnn_z3d}.   These zones are defined by regions where all stray interactions are under 50 kHz."The region marked within the dashed green boundary in the final column presents zones exhibiting minimal stray couplings, both two-body and three-body. Upon comparison with scenarios of hard decoupling, it becomes evident that in the example of larger direct qubit-qubit coupling $g_{12}$, the zone of minimal stray interaction—termed the `stray safe zone'—extends over a broader area in conditions of soft decoupling than in those of hard decoupling. This suggests that enhances the practical applicability of soft decoupling techniques in managing stray couplings.

With increasing the direct coupling strength $g_{12}$, the stray safe zones becomes noteworthy wider. This suggests that a non-zero direct coupling between $ Q1 $ and $ Q2 $ can better help to keep stray coupling strength small as it may supply a broader range of operational parameters.  Similar to the previous section, in the soft isolated circuits shown Fig.~\ref{fig:fnn_z3d} we demonstrate the  $ZZ$ gate on $Q2$ and $Q3$ can be turned  on and off by tuning the coupler frequency between them, i.e. $\omega_{c23}$; i.e. this gate can be turned ON (OFF) by increasing (decreasing) $\omega_{c23}$.

\begin{figure}
\centering
\includegraphics[width=0.87\linewidth]{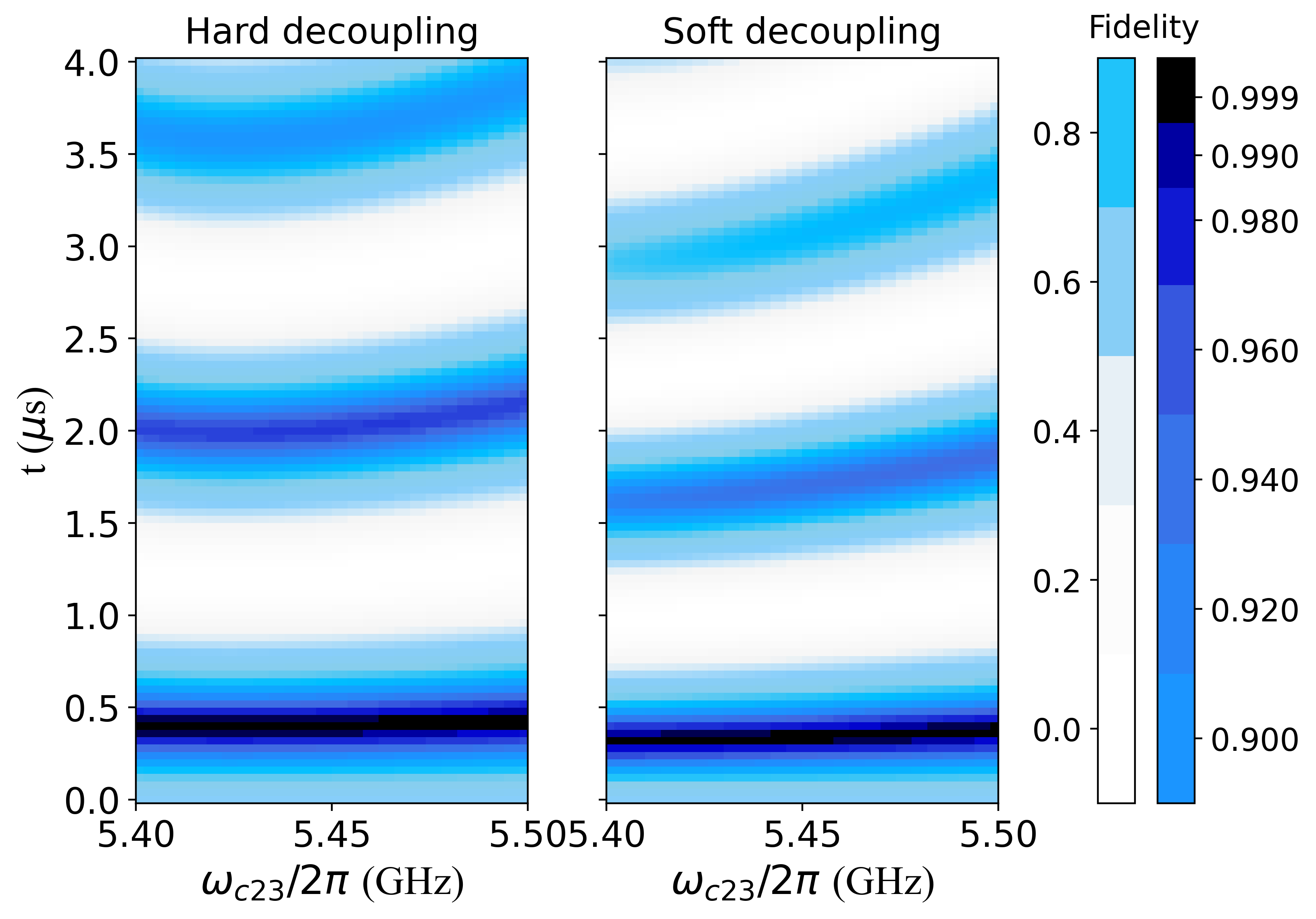}
\put(-230,142){\textbf {(a)}}\\
\includegraphics[width=0.85\linewidth]{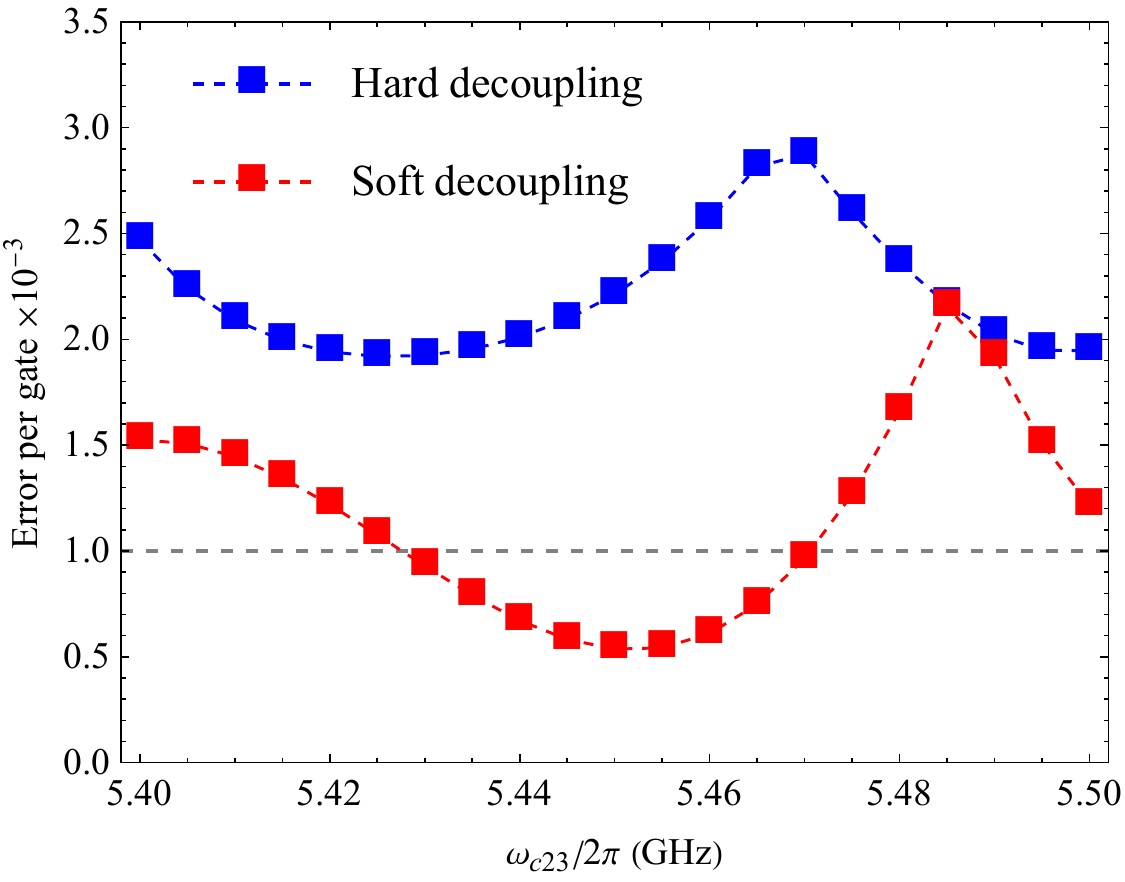}
\put(-230,155){\textbf {(b)}}
\vspace{-0.1in}
 \caption{(a) $CZ$ gate fidelity between Q2 and Q3 for the cases of hard  and soft decoupling, as a function of $\omega_{c23} $ and time $t$. (b) Maximum $CZ$ gate fidelity as a function of $\omega_{c23} $ for the cases of hard decoupling (blue) and soft decoupling (red).}
    \label{fig:fidelity}
\end{figure}

To quantify the impact of three-body $ZZZ$ interactions on two-qubit gates, we evaluate the fidelity of the $CZ$ gate between Q2 and Q3 for both hard $J$ and soft $J$ decoupling scenarios. For simplicity, we assume no decoherence from the qubits to the environment and calculate the maximum fidelity as a function of the coupler frequencies $\omega_{c23}$ at specific coupler frequency $\omega_{c13}$ within the low-error zone. The comparison in Fig.~\ref{fig:fidelity} indicates that the $ZZZ$ interaction introduces approximately a 0.2\% error, thus excluding the hard decoupling  case beyond the error correction threshold.

Overall, our comparative analysis with hard decoupling demonstrates that the soft decoupling scenario can provide equivalent or broader stray-free zones compared to the hard decoupling scenario. The comparison indicates that higher-level interactions between gated and ungated qubits play a crucial role and require careful management. These characteristics can be encapsulated in the form of $ZZZ$ interactions. In quantum processors, where the fundamental operations are single- and two-qubit gates, these interactions should be carefully suppressed either at the design stage by reducing connectivity or lowering capacitive coupling, or at the gate stage by employing the soft decoupling  method described above, which involves tuning circuit parameters such as the coupler frequency to optimize performance and suppress unwanted interactions. By integrating these strategies, we can enhance the reliability and efficiency of quantum processors.

\section{$ZZZ$ Superiority}\label{sec:three}

We study the two-body and three-body stray  interactions in superconducting quantum processors, with a particular focus on scenarios where the strength of three-body interactions $ZZZ$ surpasses that of two-body interactions. This observation would potentially affect the design consideration of future superconducting transmon architectures.

\begin{figure}[ht]
\centering
\includegraphics[width=0.9\linewidth]{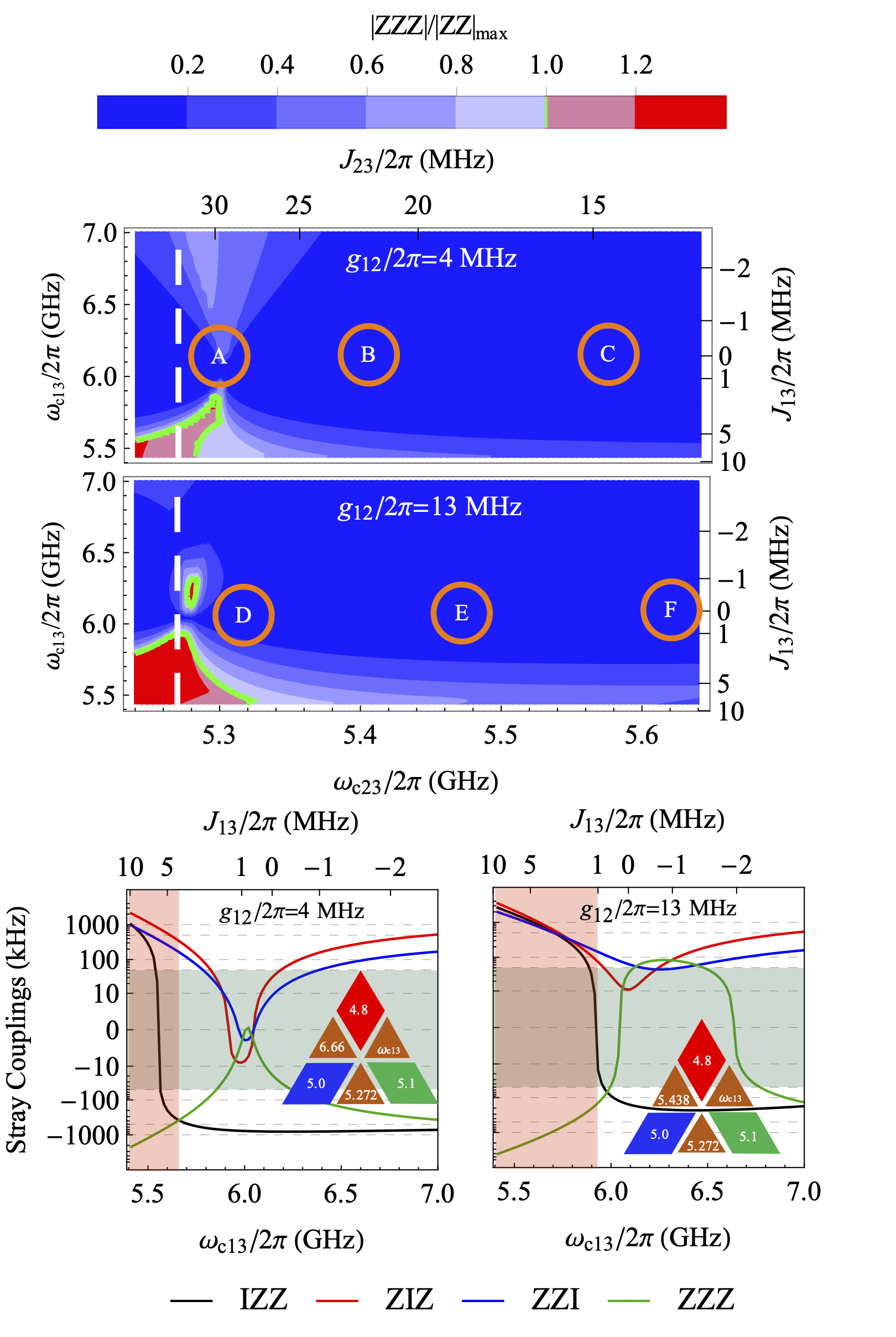}
\put(-220,280){\textbf {(a)}}
\put(-220,210){\textbf {(b)}}
\put(-220,120){\textbf {(c)}}
\put(-115,120){\textbf {(d)}}
\vspace{-0.1in}
\caption{Absolute ratio of the three-body $ZZZ$ and the maximum of the two-body interactions occurring in the triangular circuit in Fig.~\ref{fig:3qubit} as a function of the frequencies of resonators C23 and C13 $\omega_{c23}$ and $\omega_{c13}$ at (a) $g_{12}/2\pi=4$ MHz and (b) $g_{12}/2\pi=13$ MHz, respectively. The green contour line represents $|\alpha_{ZZZ}|={\rm max}(|\alpha_{Z_iZ_jI_k}|)$. Corresponding stray couplings at  specific coupler frequency $\omega_{c23    }/2\pi=5.272$ GHz (dashed white line on the top plots) are shown in (c) and (d), respectively. } Gray area indicate stray couplings are less than 50 kHz, light green area show the validity of $ZZZ$ superiority. Other circuit parameters are the same as in Fig.~\ref{fig:zzzvszz} with $\omega_2/2\pi=5.0$ GHz.
\label{fig:ffn_zratio}
\end{figure}
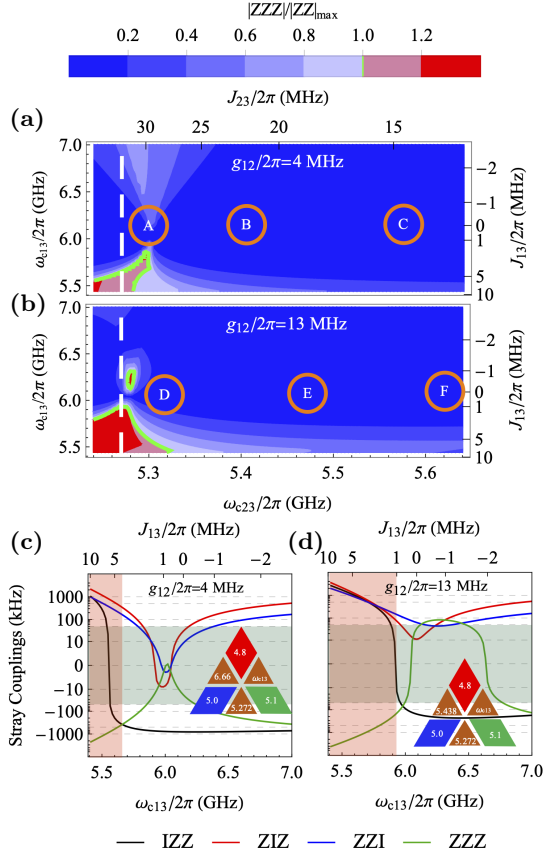

In the circuits described in the section of hard decoupling , we parameterize the circuits so that $Q1$ does not $J^{00}$-interact with $Q2$ and $Q3$. We evaluated $ZZZ$ and $ZZ$ strengths in section \ref{sec. strict}.  We use the data and compute the ratio of $|ZZZ|$ to the maximum two-body interaction strength under different capacitive coupling strengths between $Q1$ and $Q2$, e.g.  $g_{12}/2\pi=4,\ 13$~MHz. The results are depicted in Fig.~\ref{fig:ffn_zratio}(a,b).   While the marked blue region aligns with the expected hierarchies in MBL theory, the red areas reveal specific domain of parameters under which three-body interactions exceed two-body interactions, particularly at coupler frequencies close to the qubit frequencies.  In these regions  $ZZZ$ interactions become more dominant than two-body interactions for certain coupler frequencies, namely `$ZZZ$ superiority.' In Fig.~\ref{fig:ffn_zratio}, the `$ZZZ$ superiority' region is highlighted, with its boundary indicated by a solid green line where $|ZZZ|=|ZZ_{\max}|$. The increase in the prominence of three-body interactions with stronger direct coupling is a noteworthy observation.

Circles labeled A to F in Fig.~\ref{fig:ffn_zratio}(a,b) represent three sweet spots for the OFF-OFF-ON gate studied in Fig.~\ref{fig:ffn_z2d}. The analysis indicates that in certain ON spots  which are near the three-body superiority zone, such as point A, are not ideal as a safe gate sweet spot. In such operational points, while $ZZ$ strengths are small, the three-body  $ZZZ$ interaction can exceed $ZZ$'s, which can lead to noisy gate operations. Figure~\ref{fig:ffn_zratio} (c, d) show the behavior of stray couplings at the fixed frequency $\omega_{C_{23}}/2\pi=5.272$ GHz, represented by a white dashed line in Fig.~\ref{fig:ffn_zratio} (a, b). The behavior is examined across a range of $\omega_{C_{23}}$ frequencies for both weak and strong $g_{12}$.  In the weak direct coupling $g_{12}$ case, a narrow band around $\omega_{c13}/2\pi=6$ GHz can be identified in which all residual stray interactions are small. This observation indicates that the specified frequency range may act as a zone with minimal error, optimal for establishing a quantum idle point. Consequently, this parameter domain is suitable for implementing a Controlled-$Z$ ($CZ$) gate between qubits Q2 and Q3. Another example of a three-qubit circuit, where three qubits are coupled to a shared coupler, is provided in Appendix F.

This analysis highlights the intricate balance of multi-qubit interactions in quantum circuits, emphasizing the need to consider both two-body and three-body interactions in circuit design. Especially in scenarios aiming for high-fidelity gate operations, the findings suggest that a nuanced understanding of these interactions can lead to optimized strategies for quantum computing. This approach avoids oversimplified assumptions of interaction hierarchies and paves the way for more effective quantum gate designs.

\section{Cross Resonance gate on Lattice}\label{app:crgate}

Cross resonance ($CR$) gate is a type of interacting entanglement between two qubits, applied locally on one qubit, namely `control', and makes the other qubit, namely `target', undergoes $\pm \hat{X}$ operation, with $\pm$ being subject to the state of the control qubit; i.e. the interacting Hamiltonian between is $ZX$. This can be done in the lab by driving the control qubit with a microwave pulse of frequency of the target qubit~\cite{paraoanu06microwave-induced,rigetti10fully}. Here we study the impact of microwave driving on the strength of stray couplings.
 
Applying $CR$ gate on qubits makes an impact on the stray couplings between qubits and this has been previously extensively studied in \cite{xu2020high-fidelity,xu2021zz-freedom,kandala21demonstration,xu2023parasitic-free}, even in Ref. \cite{xu2021zz-freedom,ansari0circuit,ansari0method} it has been proposed that the microwave activated part of the stray coupling could take place with the opposite sign of the static stray coupling of no $CR$ gate, so that by tuning circuit parameters one can cancel the two and make the so-called $ZZ$-free qubits, which is yet to be verified experimentally.  However, detailed impact of $CR$ gate on the lattice Hamilton in many-body form is unknown. 

In this section study the impact of $CR$ gate on $ZZ$ and $ZZZ$ for a system of superconducting qubits. The microwave driving Hamiltonian in the lab frame is $H_{\textup{drive}}=\Omega \cos (\omega_{\textup{target}}t) (\hat{a}_{\textup{control}}+\hat{a}^\dagger_{\textup{control}})$. We take this part of Hamiltonian into the relevant frame of  non-perturbative block diagonalization, and together with the static Hamiltonian in the basis of $|Q_{\textup{control}},  Q_{\textup{target}}, Q_{\textup{spectator}} \rangle$, final result will end up in the following general form: 
\begin{eqnarray}
\label{eq.H CRfull}
H&=&\alpha_{ZZI} \frac{ZZI}{4} + \alpha_{ZIZ} \frac{ZIZ}{4} + \alpha_{IZZ} \frac{IZZ}{4} +\alpha_{ZZZ} \frac{ZZZ}{8} \nonumber \\
&& + \alpha_{ZXI} \frac{ZXI}{2} + \alpha_{ZXZ} \frac{ZXZ}{4}     
\end{eqnarray}

Here classical crosstalk is assumed to be canceled. The two-qubit version of these Pauli coefficients ---$\alpha_{ijk}$ with one of the three indices being identity and the other two from the set $\{I, Z, X, Y \}$--- have been analytically determined by perturbative block diagonalization techniques in \cite{xu2021zz-freedom,magesan2020effective}. Non-perturbative numerical analysis show that the pairwise Pauli coefficients in the leading order of CR gate amplitude scale as $\alpha_{ZZ}(\Omega)=\alpha_{ZZ}(0)+\eta_2 \Omega^2+\eta_a \Omega^a$ with coupler frequency dependent power $4\leq a \leq 5$, and $\alpha_{ZX}(\Omega)=\mu_1 \Omega + \mu_b \Omega^b$ with $b$ being nearly 3. For more discussion on the method of evaluation and domain of parameters see Ref.~\cite{xu2023parasitic-free}, Fig. 8.  It is evident that $ZZZ$ under the microwave drive of a $CR$ gate changes from its static value by the following form, which can be verified perturbatively: $\alpha_{ZZZ}(\Omega)=\alpha_{ZZZ}(0)+\nu_{2} \Omega^2+O(3)$ in the leading order, with $\nu_2$ depending on the parameters of circuit and can be positive or negative, meaning that the microwave-activated part of the $ZZZ$ can add or reduce error from the circuit lattice Hamiltonian.

\begin{figure}[h!]
\centering
\includegraphics[width=\linewidth]{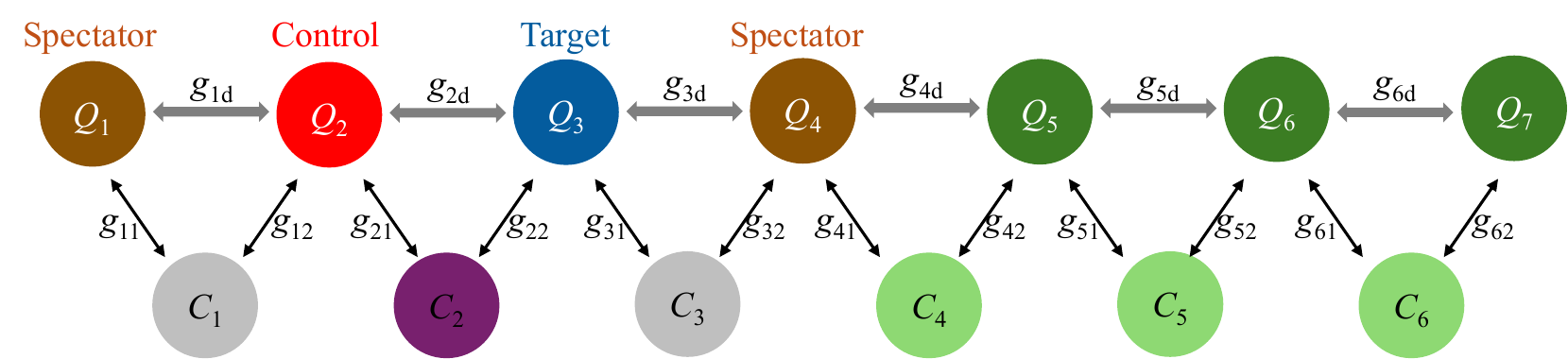}
 \caption{Schematic of thirteen Xmon qubits with the bottom six qubits serving as tunable couplers. See Ref.~\cite{cai2021impact} for more details.}
    \label{fig:7q}
\end{figure}

\begin{figure*}[t]
\centering
\includegraphics[width=0.75\linewidth]{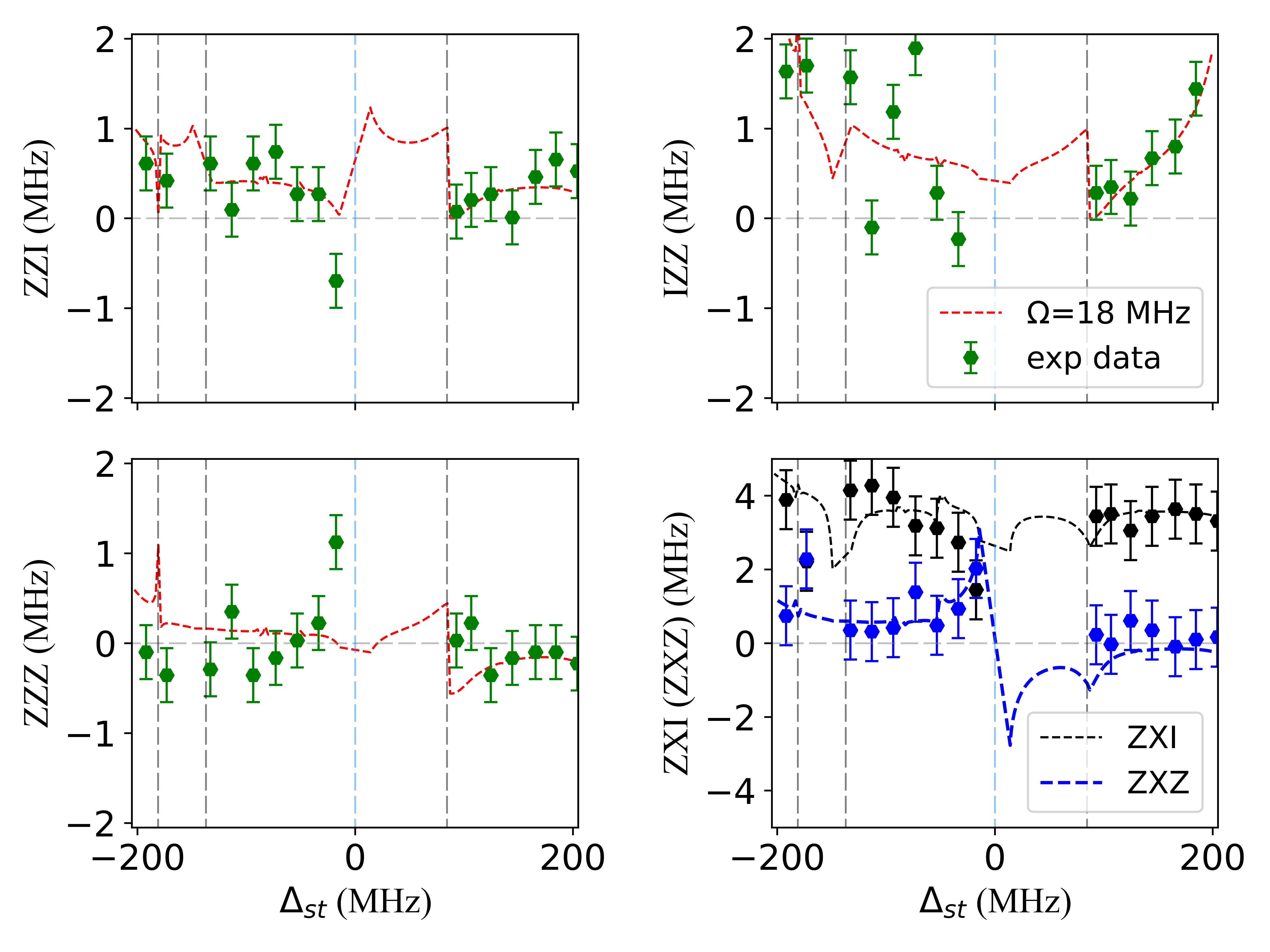}
\put(-390,275){\textbf {(a)}}
\put(-195,275){\textbf {(b)}}
\put(-390,150){\textbf {(c)}}
\put(-195,150){\textbf {(d)}}
\vspace{-0.15in}
 \caption{Interaction terms as varying the spectator-target detuning $\Delta_{st}$ with fixed CR drive amplitude $\Omega/2\pi=18$ MHz when
considering parasitic interactions from the target spectator qubit: (a) $ZZI$, (b) $IZZ$, (c) $ZZZ$ and (d) $ZXI$ ($ZXZ$). Markers represent the extracted experimental data in Ref.~\cite{cai2021impact}, curves are the numerical simulation results.}
    \label{fig:repexp}
\end{figure*}
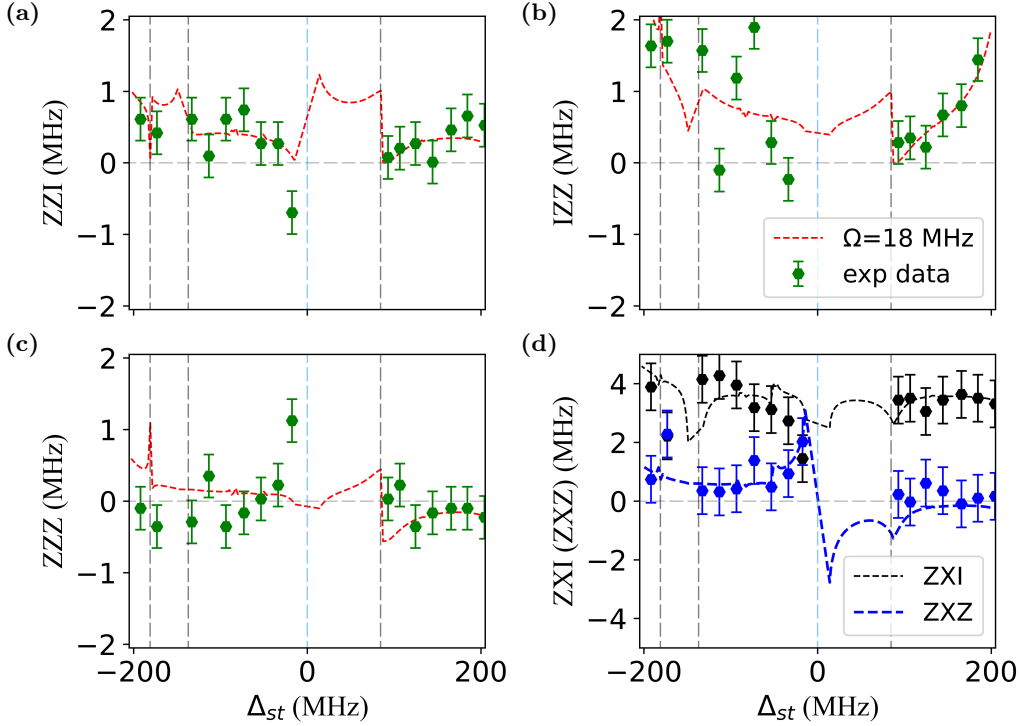

We evaluate the Pauli coefficients in the Lattice Hamiltonian of Eq.~(\ref{eq.H CRfull}) numerically. In order to compare our simulation results with experiment we study the recent measurement data of a 7-qubit circuit  in Ref.~\cite{cai2021impact}, as shown in Fig.~\ref{fig:7q}. Qubits interact by cross-resonance gate. In particular, the authors assess spectator errors through three-qubit tomography in Fig.~S7(D). In this case, let \(Q_2\) be the control qubit, \(Q_3\) the target, and \(Q_4\) the target spectator qubit, with no coupler between \(Q_2\) and \(Q_4\). Here, we employ our non-perturbative approach to simulate the corresponding Pauli coefficients at the driving amplitude $\Omega/2\pi=18$ MHz, illustrated in Fig.~\ref{fig:repexp}. In order to match the measured $ZZZ$ and $ZXZ$ data, we assume \(Q_2\) and \(Q_4\) are capacitively coupled with a strength of $g_{24d}/2\pi=4.5$ MHz. Markers depict the extracted experimental data, while curves represent results from our numerical method. Our theory effectively predicts the behavior of such qubit chain devices.

Table~\ref{tab:resonance} summarizes several static and microwave-activated resonances ---perturbative divergences--- caused in the presence of a third qubit in the circuit, as discussed in Ref.~\cite{cai2021impact}. The static (microwave-activated) resonances are highlighted in vertical blue (black) dashed lines  in Fig.~\ref{fig:zzzvszz}. The introduction of a third qubit leads to numerous higher-level divergences that are not accounted for in the two-qubit analysis.

\begin{table}[ht]
\label{tab.div}
\begin{center}
\begin{tabular}{ | c | c| c |}
	\hline 
	\scriptsize{List of static resonances} &\scriptsize{Condition} &\scriptsize{$\frac{\Delta_{st}}{2\pi}\textup{[MHz]}$}
		\tabularnewline\hline 
	$|001\rangle\sim|010\rangle$   &\multirow{2}*{$\Delta_{st}=0$} &\multirow{2}*{$0$}\\
	$|101\rangle\sim|110\rangle$& &
			\tabularnewline\hline 
			$|001\rangle\sim|100\rangle$	&\multirow{2}*{$\Delta_{cs}=0$} &\multirow{2}*{137} \\
			$|011\rangle\sim|110\rangle$& &
			\tabularnewline\hline 			$|011\rangle\sim|020\rangle$	 &\multirow{2}*{$\Delta_{st}=\delta_2$} &\multirow{2}*{$-218$}\\
			$|111\rangle\sim|120\rangle$ & & 
			\tabularnewline\hline 			$|011\rangle\sim|002\rangle$&\multirow{2}*{$\Delta_{st}=-\delta_3$} &\multirow{2}*{$213$}\\
			$|111\rangle\sim|102\rangle$ & &
			\tabularnewline\hline 			$|011\rangle\sim|200\rangle$&$\Delta_{cs}+\Delta_{cs}=-\delta_1$ & $56$
			\tabularnewline\hline 			$|101\rangle\sim|200\rangle$ 	&\multirow{2}*{$\Delta_{cs}=-\delta_1$} &\multirow{2}*{$-81$}\\
			$|111\rangle\sim|210\rangle$ & &
			\tabularnewline\hline 			$|101\rangle\sim|002\rangle$ &\multirow{2}*{$\Delta_{cs}=\delta_3$} &\multirow{2}*{$350$}\\
			$|111\rangle\sim|012\rangle$ & &
			\tabularnewline\hline 			$|101\rangle\sim|020\rangle$&$\Delta_{ct}+\Delta_{st}=\delta_2 $ & $-355$
			\tabularnewline\hline 			$|110\rangle\sim|002\rangle$&$\Delta_{cs}-\Delta_{st}=\delta_3$ & $175$
			\tabularnewline\hline 			$|020\rangle\sim|002\rangle$&$2\Delta_{st}=\delta_2-\delta_3 $ & $-2.5$
			\tabularnewline\hline 			$|200\rangle\sim|002\rangle$&$2\Delta_{st}-2\Delta_{ct}=\delta_1-\delta_3 $ & $134.5$
			\tabularnewline\hline 			$|012\rangle\sim|120\rangle$&$2\Delta_{st}=\Delta_{ct}+\delta_2-\delta_3 $ & $66$\\
			\hline \hline
	\scriptsize{List of CR-activated resonances}&\scriptsize{Condition} & \scriptsize{$\frac{\Delta_{st}}{2\pi}\textup{[MHz]}$} 
	\tabularnewline\hline       $|000\rangle|n_d+2\rangle \sim |002\rangle|n_d\rangle$&$2\Delta_{st}=-\delta_3$ & $106.5$
	\tabularnewline\hline       $|000\rangle|n_d+2\rangle \sim |101\rangle|n_d\rangle$&$\Delta_{st}=-\Delta_{ct}$ & $-137$
	\tabularnewline\hline       $|300\rangle|n_d+1\rangle \sim |202\rangle|n_d\rangle$&$2\Delta_{st}=\Delta_{ct}+2\delta_1-\delta_3$ & $-180$
	\tabularnewline\hline       $|003\rangle|n_d+1\rangle \sim |400\rangle|n_d\rangle$&$3\Delta_{st}=4\Delta_{ct}+6\delta_1-3\delta_3$ & $-40.3$\\ \hline

        \end{tabular}
    \end{center}
	\vspace{-0.15in}
	\caption{\label{tab:resonance}Summary of three-qubit resonances in the three-qubit subspace. The states are denoted in the following order: $|Q_{\textup{control}}Q_{\textup{target}}Q_{\textup{spectator}}\rangle$. The circuit parameters are same as in Fig.~\ref{fig:zzzvszz}.}
\end{table}

In order to further our understanding of the impact of CR gate on the lattice Hamiltonian of three qubits, we use our software {\href{cirqubit.com} {CirQubit}} to generate numerical data for the cases that experiment has not conducted, such as the static stray coupling in absence of $CR$ gate $\Omega=0$, and the case of $CR$ gate amplitude being $\Omega=31$ MHz, which is almost the double of the experimental data taken at $\Omega=16$ MHz. Results have been plotted in Appendix \ref{appen.cr} Fig.~\ref{fig.CRcompare}, which indicates how the Lattice Pauli coefficients of Eq.~(\ref{eq.H CRfull}) depends on the frequency detuning between control and target qubits $\Delta_{st}$. Interestingly we see in Fig.~\ref{fig.CRcompare} in a large domain of $\Delta_{st}$, by increasing the microwave power, the interactions $ZXI$ and $ZXZ$ become more powerful, while $ZZ$ and $ZZZ$ stray couplings are suppressed.  The suppression of $ZZ$'s have been predicted earlier in Ref.~\cite{xu2021zz-freedom} known as `$ZZ$-freedom.'  Here, we extend that result by demonstrating that it is possible to tune the circuit parameters to suppress the three-body stray coupling $ZZZ$.

\section{Conclusion}
In conclusion, this study underscores the critical importance of considering the lattice Hamiltonian and many-body couplings in the design and operation of quantum circuits. This approach allows for fine-tuning circuit parameters to mitigate the detrimental effects of stray couplings. Furthermore, we demonstrate that three-body stray couplings can be as significant as two-body interactions, particularly beyond the dispersive regime. Effectively controlling these interactions through strategic adjustments can enhance gate performance and holds potential for the implementation of high-fidelity two-qubit gates.

Additionally, we have shown that the dominance of three-body $ZZZ$ interactions can be realized within the quasi-dispersive regime, suggesting the possibility of achieving optimal quantum performance even outside the near-degenerate regimes. These findings offer significant contributions to the fields of quantum system design and algorithm optimization, promising enhancements in the efficiency and robustness of quantum computing technologies.

\section*{Acknowledgement} 
 M Ansari would like to express gratitude for the technical support and enriching discussions with Jos\'ephine Pazem that greatly contributed to the development of this work. X. Xu was partially supported by the Project GeQCoS, Grant No. 13N15685. Manabputra's involvement was partially supported by the Kishore Vaigyanik Protsahan Yojana (KVPY) Fellowship, sponsored by the Department of Science \& Technology, Government of India. This research received funding from Horizon Europe (HORIZON) Project: 101113946 OpenSuperQPlus100.

\appendix

\section{Perturbative $ZZ$ and $ZZZ$ on a triangular lattice}

The explicit formulas for stray couplings, derived from perturbation theory, are presented below:

\begin{align}\alpha_{ZZI}= & \frac{2\left(J_{12}^{01}\right)^{2}}{\Delta_{12}^{01}}-\frac{2\left(J_{12}^{10}\right)^{2}}{\Delta_{12}^{10}}+\frac{4J_{12}^{00}J_{23}^{00}J_{13}^{00}}{\Delta_{23}^{00}\Delta_{13}^{00}}\nonumber \\
 & +4\Bigg(\frac{J_{12}^{00}J_{23}^{01}J_{13}^{01}}{\Delta_{23}^{01}\Delta_{13}^{01}}-\frac{J_{13}^{00}J_{12}^{01}J_{23}^{10}}{\Delta_{23}^{10}\Delta_{13}^{00}}-\frac{J_{23}^{00}J_{13}^{10}J_{12}^{10}}{\Delta_{13}^{10}\Delta_{23}^{00}}\nonumber\\
 & -\frac{J_{13}^{11}J_{23}^{01}J_{12}^{10}}{\Delta_{23}^{01}\Delta_{12}^{10}}+\frac{J_{23}^{11}J_{12}^{01}J_{13}^{01}}{\Delta_{12}^{01}\Delta_{13}^{01}}+\frac{J_{12}^{11}J_{13}^{10}J_{23}^{10}}{\Delta_{13}^{10}\Delta_{23}^{10}}\Bigg)
\end{align}
\vspace{-0.15in}
\begin{align}\alpha_{ZIZ}= & \frac{2\left(J_{13}^{01}\right)^{2}}{\Delta_{13}^{01}}-\frac{2\left(J_{13}^{10}\right)^{2}}{\Delta_{13}^{10}}+\frac{4J_{13}^{00}J_{23}^{00}J_{12}^{00}}{\Delta_{23}^{00}\Delta_{12}^{00}}\nonumber \\
 & +4\Bigg(-\frac{J_{13}^{00}J_{12}^{01}J_{23}^{10}}{\Delta_{12}^{01}\Delta_{23}^{10}}+\frac{J_{12}^{00}J_{23}^{01}J_{13}^{01}}{\Delta_{12}^{00}\Delta_{23}^{01}}+\frac{J_{23}^{00}J_{13}^{10}J_{12}^{10}}{\Delta_{23}^{00}\Delta_{12}^{10}}\nonumber \\
 & -\frac{J_{13}^{11}J_{23}^{01}J_{12}^{10}}{\Delta_{23}^{01}\Delta_{12}^{10}}+\frac{J_{23}^{11}J_{12}^{01}J_{13}^{01}}{\Delta_{12}^{01}\Delta_{13}^{01}}+\frac{J_{12}^{11}J_{13}^{10}J_{23}^{10}}{\Delta_{13}^{10}\Delta_{23}^{10}}\Bigg)
\end{align}

\vspace{-0.15in}
\begin{align}\alpha_{IZZ}= & \frac{2\left(J_{23}^{01}\right)^{2}}{\Delta_{23}^{01}}-\frac{2\left(J_{23}^{10}\right)^{2}}{\Delta_{23}^{10}}+\frac{4J_{23}^{00}J_{13}^{00}J_{12}^{00}}{\Delta_{13}^{00}\Delta_{12}^{00}}\nonumber \\
 & +4\Bigg(\frac{J_{23}^{00}J_{12}^{10}J_{13}^{10}}{\Delta_{12}^{10}\Delta_{13}^{10}}-\frac{J_{12}^{00}J_{23}^{01}J_{13}^{01}}{\Delta_{12}^{00}\Delta_{13}^{01}}-\frac{J_{13}^{00}J_{23}^{10}J_{12}^{01}}{\Delta_{13}^{00}\Delta_{12}^{10}}\nonumber \\
 & -\frac{J_{13}^{11}J_{23}^{01}J_{12}^{10}}{\Delta_{23}^{01}\Delta_{12}^{10}}+\frac{J_{23}^{11}J_{12}^{01}J_{13}^{01}}{\Delta_{12}^{01}\Delta_{13}^{01}}+\frac{J_{12}^{11}J_{13}^{10}J_{23}^{10}}{\Delta_{13}^{10}\Delta_{23}^{10}}\Bigg)
\end{align}
\vspace{-0.15in}

\begin{align}\alpha_{ZZZ}= & 8\Bigg(\frac{J_{13}^{00}J_{23}^{10}J_{12}^{01}}{\Delta_{23}^{10}\Delta_{12}^{01}}-\frac{J_{12}^{00}J_{13}^{01}J_{23}^{01}}{\Delta_{23}^{01}\Delta_{13}^{01}}-\frac{J_{23}^{00}J_{12}^{10}J_{13}^{10}}{\Delta_{12}^{10}\Delta_{13}^{10}}\nonumber \\
 & +\frac{J_{13}^{11}J_{23}^{01}J_{12}^{10}}{\Delta_{23}^{01}\Delta_{12}^{10}}-\frac{J_{23}^{11}J_{12}^{01}J_{13}^{01}}{\Delta_{12}^{01}\Delta_{13}^{01}}-\frac{J_{12}^{11}J_{13}^{10}J_{23}^{10}}{\Delta_{13}^{10}\Delta_{23}^{10}}\Bigg)
\end{align}

\section{Reduced circuit Hamiltonian}\label{app:reducedH}
In order to derive the analytical formulas mentioned in Eqs.~\eqref{eq:zz}-\eqref{eq:zz3} in the main text we start from the Hamiltonian in Eq.~\eqref{eq:zzz} which describes the circuit in Fig.~\ref{fig:3qubit} with three qubits coupled to one another by three couplers. We assume rotating-wave approximation and ignore the fast oscillating terms in the Hamiltonian to get the new Hamiltonian to be 
\begin{align}
\frac{H}{{\hbar}}=&\sum_{r=1}^3\omega_{c_r}c_r^{\dagger}c_r+
\sum^3_{q=1}\sum_{n_q}\omega_{n_q}|n_q\rangle\langle n_q|\nonumber\\
&+\sum_{r=1}^3\sum_{q=1} g_{qc_r}(c_ra_q^{\dagger}-c_r^{\dagger}a_q)\nonumber\\
&+\sum_{q\neq q'}\sum_{q'=1}^3g_{qq'}(a_qa_{q'}^{\dagger}-a_
q^{\dagger}a_{q'})
\end{align}

Using first order Schrieffer–Wolff transformation we block diagonalize the Hamiltonian to decouple the couplers from the qubits. The only qubit Hamiltonian thus obtained is

\begin{align}
\frac{H_{\rm s}}{\hbar}=&\sum^3_{q=1}\sum_{n_q}\tilde{\omega}_{n_q}|n_q\rangle\langle n_q|+\sum_{q\neq q'}\sum_{n_q,m_{q'}}\sqrt{n_q+1}
    \sqrt{m_{q'}+1}\nonumber\\
   &+ J_{qq'}^{n_qm_{q'}}(|n_q+1,m_{q'}\rangle\langle n_q,m_{q'}+1|+h.c.).
\end{align}

We then restrict this Hamiltonian to the computational subspace and fully diagonalize it using Schrieffer–Wolff transformation to the third order and rewrite it in Pauli basis as in Eq.~\eqref{eq:Hz} in the main text with the coefficients given by Eqs.~\eqref{eq:zz}-\eqref{eq:zz3}. 

\section{Qubit coupling symmetry} \label{App.Identities}
In the paper, we presented a number of identities for the effective $J$ coupling strength between qubits, Eq.~\eqref{eq:J_relation} and \eqref{eq.third}. These relations can be proven using the perturbative definition of the coupling Eq. \eqref{eq. Jeff}. 

\begin{figure}[ht]
    \centering
    \includegraphics[width=0.8\linewidth]{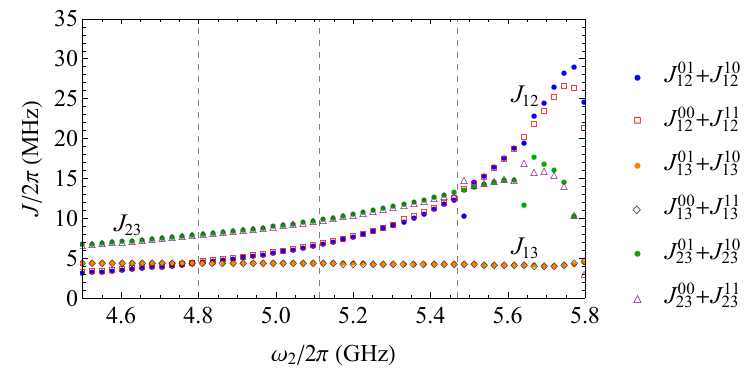}\vspace{-0.15in}
    \caption{Numerical simulation of $J$-relations in Eq.~\eqref{eq:J_relation}. The circuit parameters are same as those in Fig.~\ref{fig:zzzvszz}.}
    \label{fig:Jrel}
\end{figure}

\begin{figure}[ht]
    \centering
    \includegraphics[width=0.8\linewidth]{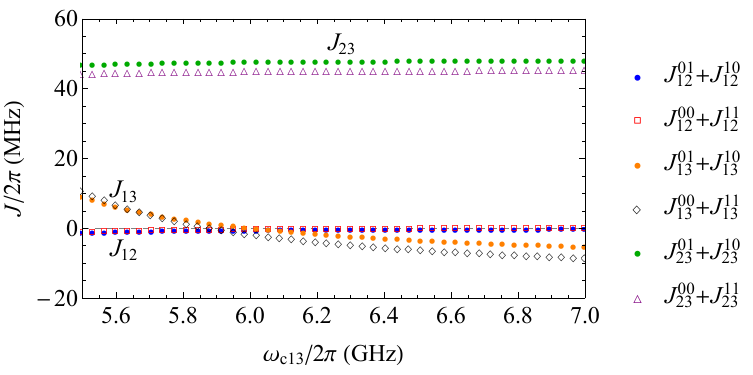}\vspace{-0.15in}
    \caption{Numerical simulation of $J$-relations in Eq.~\eqref{eq:J_relation}. The circuit parameters are same as those in Fig.~\ref{fig:ffn_zratio}(c).}
    \label{fig:Jreld}
\end{figure}

It is interesting to numerically verify these identities beyond dispersive regime as well and one way to do it is to numerically evaluate the two sides of the identities Eqs.~\eqref{eq:J_relation} and \eqref{eq.third} and verifying their overlap.   For this purpose, we simulate the $J$ relations using the nonperturbative approach with the circuit parameters similar to Fig.~\ref{fig:zzzvszz}.  The results are shown in Fig.~\ref{fig:Jrel}. We conduct a comparative analysis between the sums $J_{ij}^{01}+J_{ij}^{10}$ and $J_{ij}^{00}+J_{ij}^{11}$ to illustrate that in the dispersive regime, characterized by $J/\Delta \ll 1$, the relationship approximated by equation~\eqref{eq:J_relation} holds. However, as the frequency of an individual qubit, such as Q2, approaches the vicinity of the couplers, a noticeable divergence emerges between the two sums. Another illustrative example, employing circuit parameters identical to those in Fig.~\ref{fig:ffn_zratio}(c), demonstrates analogous results, as depicted in Fig.~\ref{fig:Jreld}.

\begin{figure}[ht]
    \centering
    \includegraphics[width=0.8\linewidth]{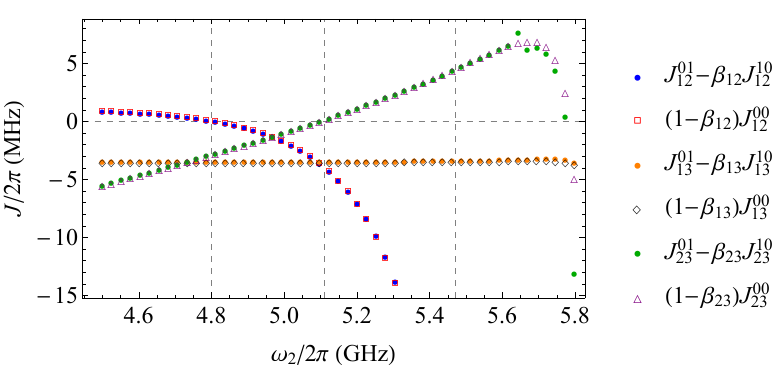}\vspace{-0.15in}
    \caption{Numerical simulation of $J$-relations in Eq.~\eqref{eq.third}. The circuit parameters are same as those in Fig.~\ref{fig:zzzvszz}.}
    \label{fig.Jre11}
\end{figure}

\begin{figure}[ht]
    \centering
    \includegraphics[width=0.8\linewidth]{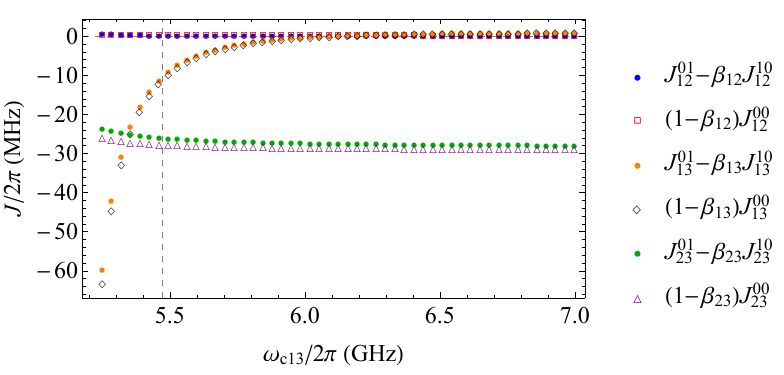} \vspace{-0.15in}
    \caption{Numerical simulation of $J$-relations in Eq.~\eqref{eq.third}. The circuit parameters are same as those in Fig.~\ref{fig:ffn_zratio}(c).}
    \label{fig.Jre11d}
\end{figure}

Figure~\ref{fig.Jre11} and \ref{fig.Jre11d} depict the numerical simulation of Eq.~\eqref{eq.third} using parameters similar to those in Fig.~\ref{fig:zzzvszz} and Fig.~\ref{fig:ffn_zratio}(c), respectively. These results exhibit significant consistency, affirming the validity of the identity.

\section{{\it ZIZ} Discrepancy}\label{app:ZIZD}

\begin{figure}[ht]
    \centering
    \includegraphics[width=0.8\linewidth]{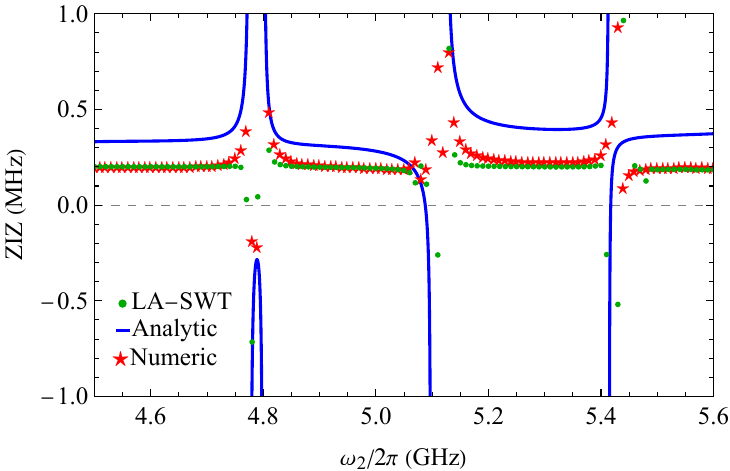}
    \caption{LA-SWT approach vs the analytical formula when compared to the numerical result. The green dots represent the LA-SWT result. The circuit parameters are same as in Fig.~\ref{fig:zzzvszz}.}
    \label{fig:zizla}
\end{figure}

Instead of relying on a perturbative block-diagonalization technique such as the Schrieffer-Wolff transformation, which is primarily accurate in the dispersive regime, an alternative approach can be employed to achieve exact multi-block-diagonalization, when analytical expressions are not required. This technique is called the least action method~\cite{cederbaum1989block,magesan2020effective} which aims to identify the block-diagonal Hamiltonian that exhibits the highest degree of similarity to the true Hamiltonian, as governed by the principle of least action. By employing this precise approach to decouple the couplers from the qubits, our analysis encompasses the higher-order interactions, as it is an exact methodology. Subsequently, the Schrieffer-Wolff transformation can be utilized to fully diagonalize the qubit Hamiltonian. Upon comparing this approach, as shown in Fig.~\ref{fig:zizla}, with the $ZIZ$ interaction depicted in Fig.~\ref{fig:zzzvszz}(b) in the main text, it becomes evident that the former exhibits greater accuracy when compared to the numerical results.

This observation indicates that the discrepancy in the $ZIZ$ interaction arises primarily from the initial Schrieffer-Wolff transformation. This is likely attributed to the fact that the effective interaction strength between two qubits, denoted as $J_{ij}^{mn}$ and derived using SWT, does not incorporate any higher order contributions from the third qubit in the circuit. However, these effects are accurately captured through the numerical simulation and the least action method.

\section{Effective Coupling Discrepancy}\label{app:Jdis}

\begin{figure}[ht]
    \centering
    \includegraphics[width=0.9\linewidth]{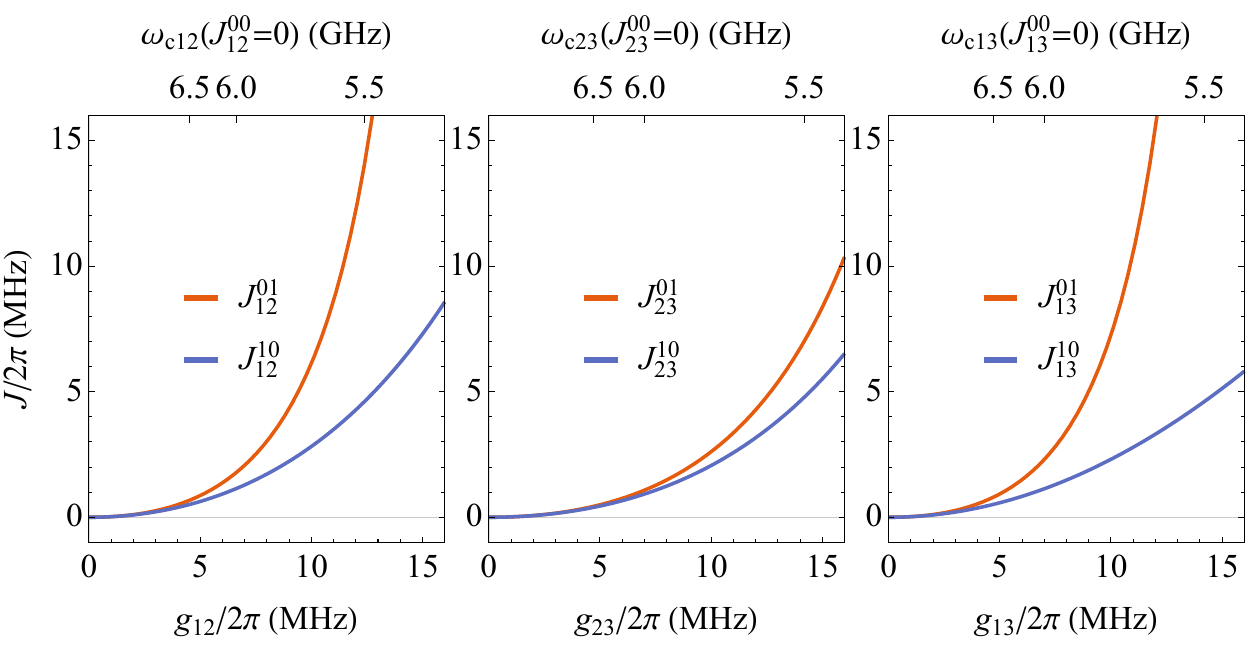}
    \put(-211,104){(a)}
    \put(-140,104){(b)}
    \put(-70,104){(c)}
    \caption{Effective qubit-qubit coupling discrepancy from analytical simulation. (a) $J_{12}^{01}$ and $J_{12}^{10}$ as a function of direct coupling $g_{12}$ when $J_{12}^{00}$ is off. (b) $J_{13}^{01}$ and $J_{13}^{10}$ as a function of direct coupling $g_{13}$ when $J_{13}^{00}$ is off. (c) $J_{23}^{01}$ and $J_{23}^{10}$ as a function of direct coupling $g_{23}$ when $J_{23}^{00}$ is off. Corresponding coupler frequency is labelled on the top.}
    \label{fig:goff}
\end{figure}

Even when the effective qubit-qubit coupling strength at the lowest order $J_{12}^{00}$ is zero, the qubits are still not necessarily fully decoupled. We can see the effect of this in Fig.~\ref{fig:ffn_z2d}. Because of the non-linearity of the superconducting qubits higher order interactions may still exist. Figure~\ref{fig:goff} depicts how the higher order interaction strength $J^{01}$ and $J^{10}$ deviate from zero by several megahertz as the direct coupling between the qubits gets stronger or when the couplers are tuned closer to the qubits. As a result, parasitic interactions can still exist between these qubits and need to be accounted for.  

\begin{figure*}[b]
\centering
\includegraphics[width=0.87\linewidth]{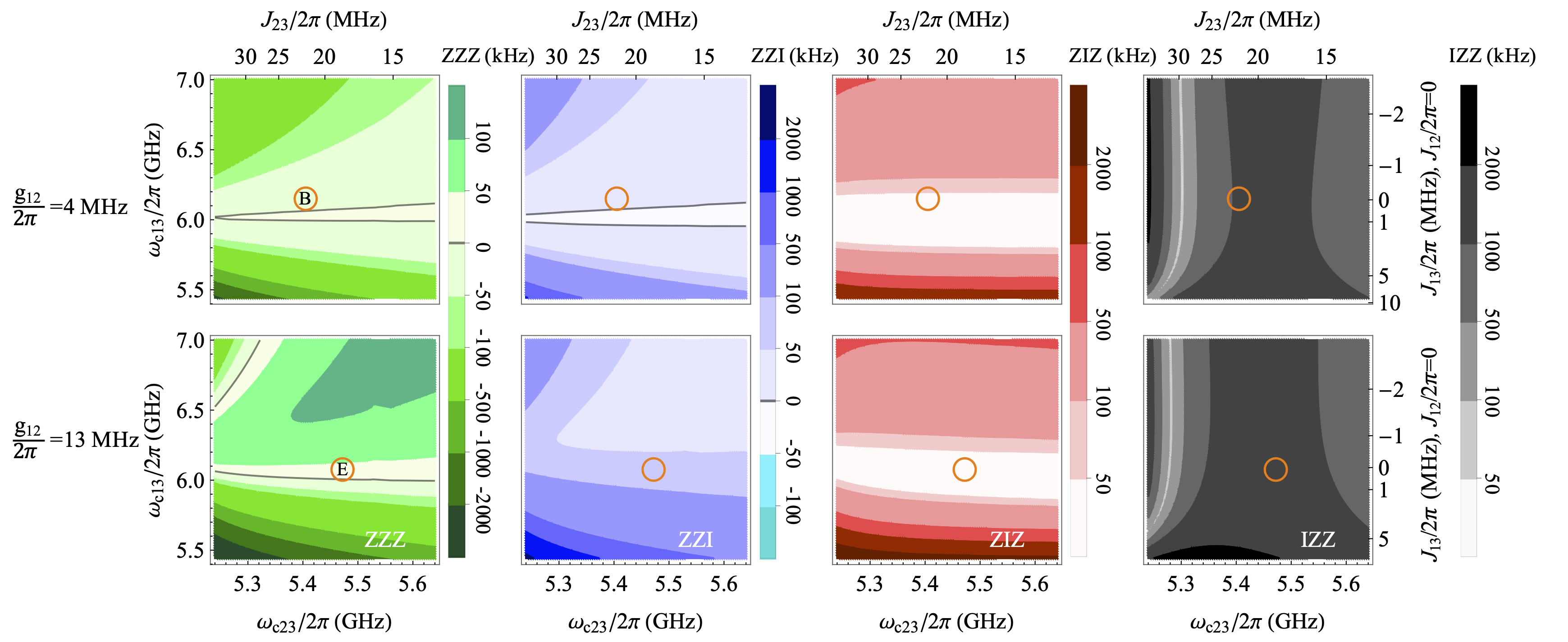}
\put(-430,170){\textbf {(a)}}\\
\includegraphics[width=0.87\linewidth]{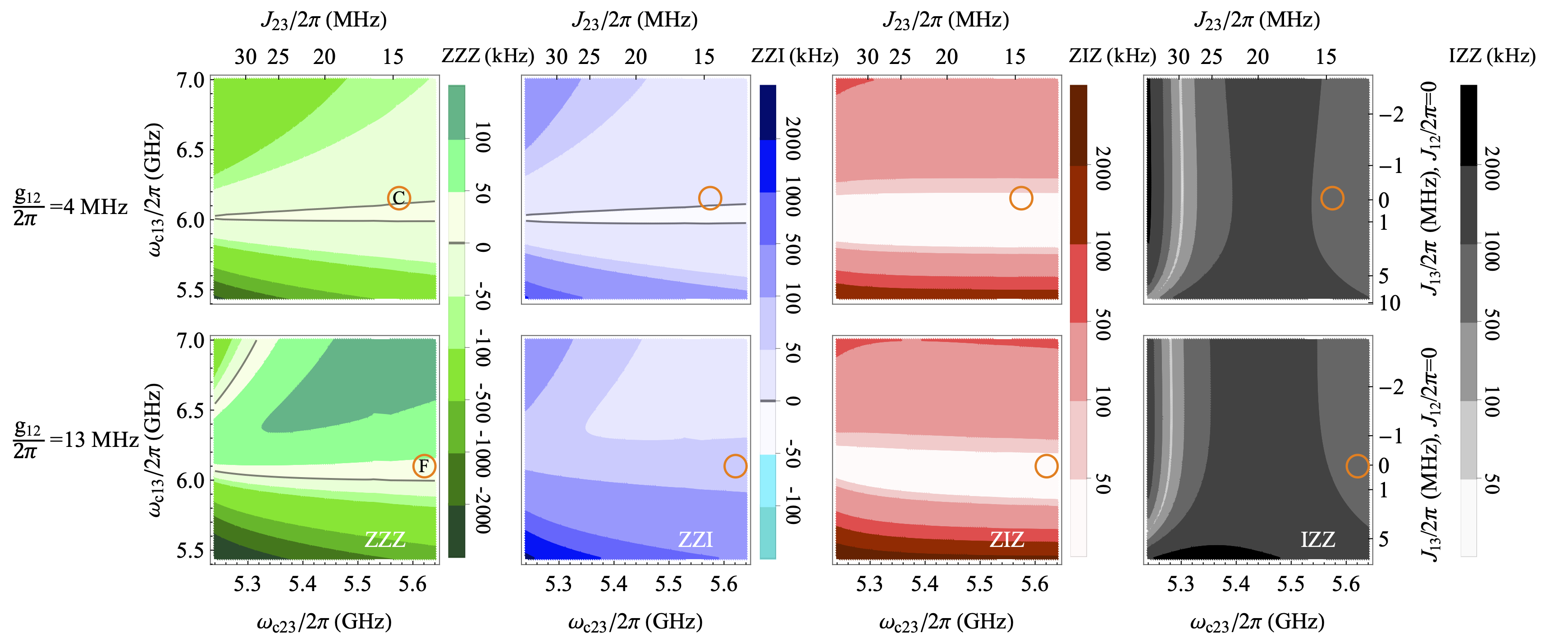}
\put(-430,170){\textbf {(b)}}
\vspace{-0.15in}
\caption{(a) Stray couplings in the triangular circuit with all-to-all connectivity as a function of the frequencies of the resonators C13 and C23 $\omega_{c13}$ and $\omega_{c23}$ respectively. The first row is at  $g_{12}/2\pi=4$ MHz with nearly zero $J_{12}$ and $J_{13}$ at spot B: $(\omega_{c12}/2\pi,\omega_{c23}/2\pi,\omega_{c13}/2\pi)=(6.67,5.405,6.15)$ GHz. The second row is at $g_{12}/2\pi=13$ MHz with nearly zero $J_{12}$ and $J_{13}$ at spot E: $(5.443,5.472,6.076)$ GHz. (b) The third (fourth) row is at  $g_{12}/2\pi=4$ (13) MHz with nearly zero $J_{12}$ and $J_{13}$ at spot C:  $(\omega_{c12}/2\pi,\omega_{c23}/2\pi,\omega_{c13}/2\pi)=(6.68,5.275,6.154)$ GHz. In the fourth row $J_{12}$ and $J_{13}$ are nearly zero at spot F: $(5.446,5.621,6.098)$ GHz. Other circuit parameters are the same as Fig.~\ref{fig:zzzvszz} with $\omega_2/2\pi=5.0$ GHz.}
    \label{fig:ffn_be}
\end{figure*}

In Fig.~\ref{fig:ffn_be}(a), we describe the landscape with C12 frequency at decoupling points at B and E. In Fig.~\ref{fig:ffn_be}(b), we describe the landscape with C12 frequency at decoupling points at C and F. Because of the higher order $J^{01}$ and $J^{10}$ interactions, the landscapes have significant features. In the case of the $ZZZ$ interaction, especially when the qubits are strongly coupled, it can go upwards of thousand kilohertz as $\omega_{c23}$ varies, even when $J_{12}$ is zero. $IZZ$ also varies significantly with $\omega_{c23}$ but $ZIZ$ have little dependence on it. In the second row of graphs, when the direct coupling $g_{12}$ is stronger, the zeroness of $ZZI$ disappears and becomes narrower for $ZIZ$.  At the decoupling points, $ZZZ$, $ZZI$ and $ZIZ$ are all smaller than 50 kHz whereas $IZZ$ is significantly stronger.       

\begin{figure*}[ht]
\centering
\includegraphics[width=0.8\linewidth]{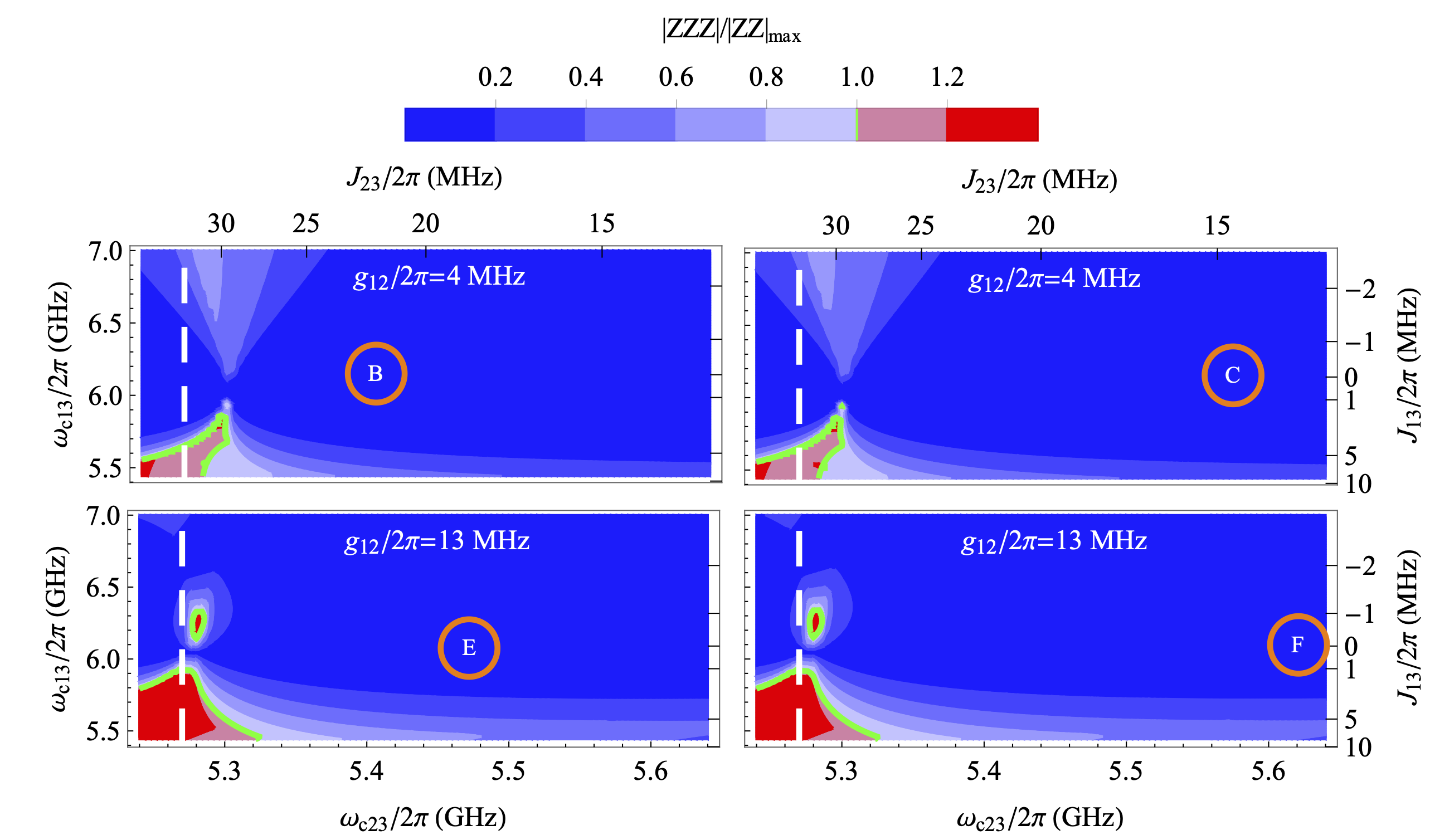}\vspace{-0.15in}
\caption{Absolute ratio of the three-body $ZZZ$ and the maximum of the two-body interactions occurring in the triangular circuit as a function of the frequencies of resonators C23 and C13 $\omega_{c23}$ and $\omega_{c13}$ at $g_{12}/2\pi=4$ MHz with spot B and C in the first two and $g_{12}/2\pi=13$ MHz with spot E and F in the second row, respectively. Other circuit parameters are the same as Fig.~\ref{fig:zzzvszz} with $\omega_2/2\pi=5.0$ GHz.}
    \label{fig:ffn_bcef}
\end{figure*}

In Fig.~\ref{fig:ffn_bcef}, we plot the ratio of the $ZZZ$ interaction and the maximum of $ZZ$ interactions as we vary the coupler frequencies. The figures on the left have C12 frequency at decoupling points at B and E, whereas, the figures on the right have C12 frequency at decoupling points at C and F. We see that the $ZZZ$ interaction is stronger than all the $ZZ$ interactions when the couplers are tuned closer to the qubits. 

\section{Three qubits and one coupler}

\begin{figure}[ht]
     \includegraphics[width=0.6\linewidth]{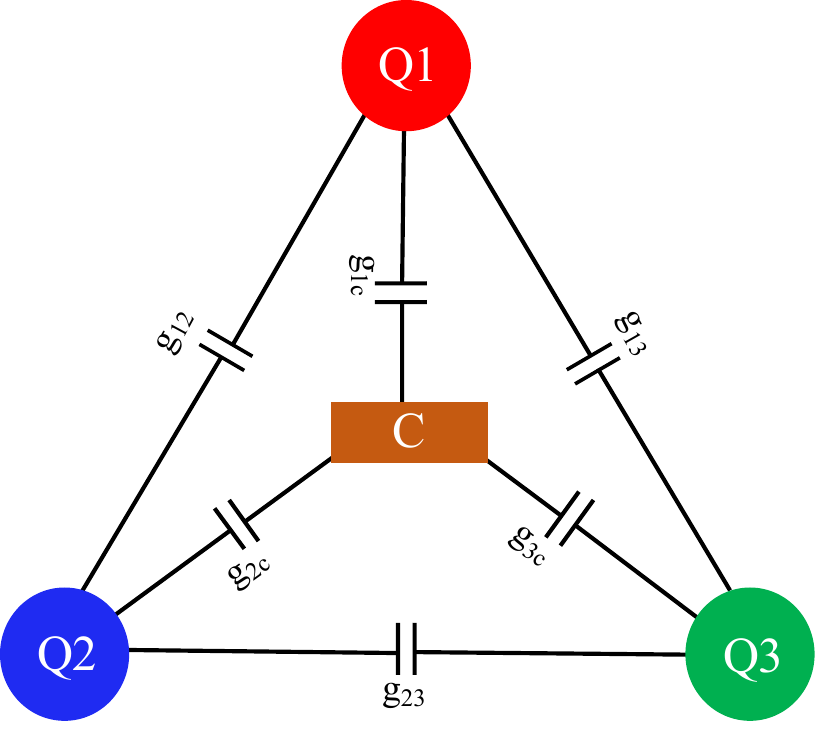}
     \put(-190,120){\textbf {(a)}}\\
     \includegraphics[width=1\linewidth]{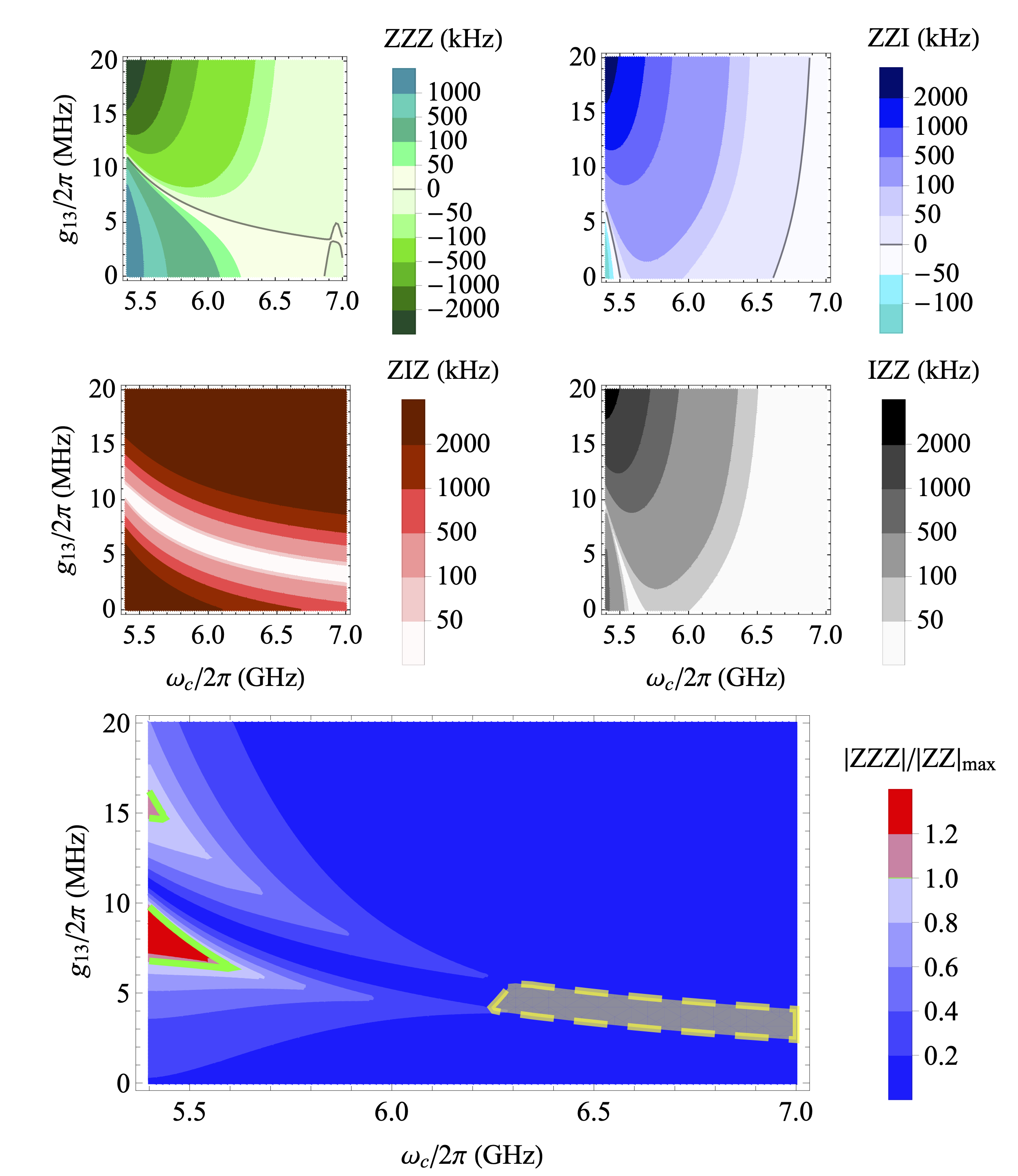}
\put(-240,270){\textbf {(b)}}
\put(-125,270){\textbf {(c)}}
\put(-240,190){\textbf {(d)}}
\put(-125,190){\textbf {(e)}}    
\put(-240,110){\textbf {(f)}}  

    \caption{(a)Triangle geometry of 3 qubit interaction indirectly coupled via one shared coupler. Stray couplings versus coupler C frequency of circuit (b) $ZZZ$, (c) $ZZI$, (d) $ZIZ$, (e) $IZZ$, (f) $|ZZZ|/{|ZZ|_{\max}}$. Circuit parameters are: $\omega_1/2\pi=4.8$ GHz, $\omega_2/2\pi=5.0$ GHz, $\omega_3/2\pi=5.1$ GHz, $\delta_1/2\pi=\delta_2/2\pi=\delta_3/2\pi=-330$ MHz with qubit-coupler strength $g_{1c1}/2\pi=g_{1c_3}/2\pi=g_{2c_1}/2\pi=g_{3c_3}/2\pi=85$ MHz and $g_{2c_2}/2\pi=g_{3c_2}/2\pi=102$ MHz, direct qubit-qubit coupling $g_{12}=g_{23}/2\pi=4$ MHz.}
    \label{fig:zzzgzz_o}
\end{figure}

We extend our investigation beyond the case of three-qubit three-coupler architecture. In this section, we study the case of three qubits sharing a coupler, which resembles the one recently studied in Ref.  \cite{menke2022demonstration}. For simplicity, we consider three transmon qubits coupled to a harmonic oscillator as the shared coupler, as depicted in Fig.~\ref{fig:zzzgzz_o}(a). Our study aims to examine the behavior of such a system in response to changes in coupler frequency $\omega_c$ and direct coupling, i.e. $g_{13}$. We present the findings in Fig.~\ref{fig:zzzgzz_o}.

Compared to the triangular circuit depicted in Fig.~\ref{fig:3qubit}, our analysis reveals that when the coupler frequency is detuned far away from the qubits, all four stray couplings are effectively suppressed to below 50 kHz at a certain direct coupling $g_{13}$. As expected, the $ZIZ$ coupling is the most influenced by $g_{13}$, while all other couplings remain almost constant at high coupler frequency, and undergo significant magnitude changes at weak coupler frequencies, where the qubit basis becomes hybridized with the coupler subspace. Additionally, we evaluate the ratio of $ZZZ$ to the maximum $ZZ$ and observe that $ZZZ$ superiority can still be achieved at lower coupler frequencies.

\section{Cross Resonance gate simulation}

\label{appen.cr}

\begin{figure*}[htbp]
\begin{center}
\includegraphics[width=0.8\linewidth]{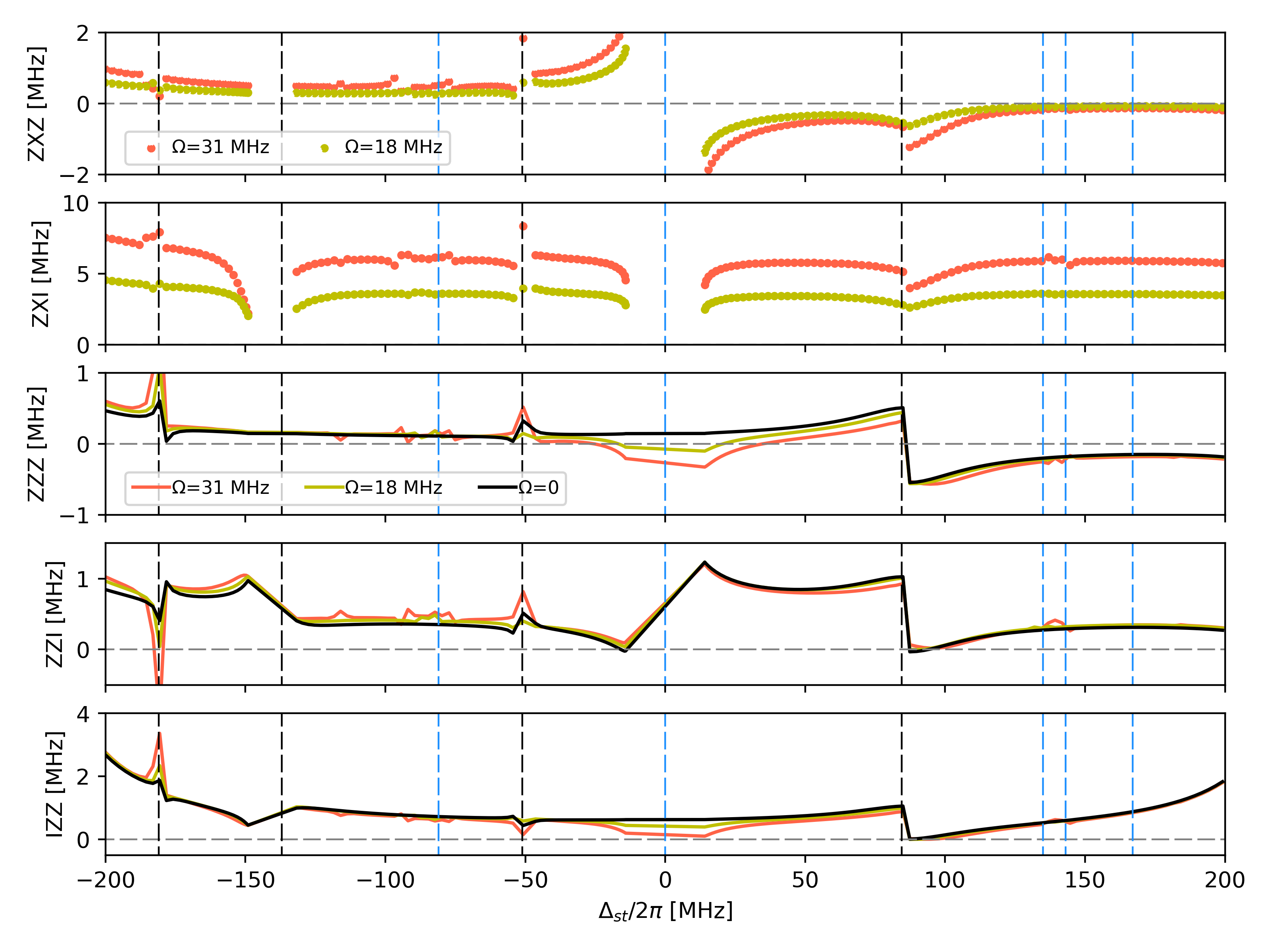}
\caption{Pauli $CR$ coefficients at three different driving amplitude: $\Omega=0$ (black), $\Omega=18$ MHz (yellow) and $\Omega=31$ MHz (pink), respectively.}
\label{fig.CRcompare}
\end{center}
\end{figure*}

We study the impact  $CR$ gate on the lattice Hamiltonian of three qubits shown in Fig.~\ref{fig:7q}. For this aim we use our software {\href{cirqubit.com} {CirQubit}} to generate numerical data for the cases that experiment has not conducted, such as the static stray coupling in absence of $CR$ gate $\Omega=0$, and the case of $CR$ gate amplitude being $\Omega=31$ MHz, which is almost the double of the experimental data taken at $\Omega=16$ MHz. Results have been plotted in Fig.~\ref{fig.CRcompare}, which indicates how the Lattice Pauli coefficients of Eq.~(\ref{eq.H CRfull}) depends on the frequency detuning between control and target qubits $\Delta_{st}$. Black dashed lines indicates static divergences that exist in the circuit in the static undriven circuit. In contrast, the blue dashed lines are new divergences that appear in the microwave-activated part.

\bibliography{ref}

\end{document}